\documentclass[nofootinbib,aps,floatfix,superscriptaddress]{revtex4}
\usepackage{tikz}
\usetikzlibrary{arrows.meta}
\usetikzlibrary{decorations.pathmorphing}

\usepackage{xcolor}
\usepackage[
    colorlinks=true,
    citecolor=blue,
    linkcolor=black,
    urlcolor=blue
]{hyperref}
\usepackage[normalem]{ulem}
\usepackage{amssymb,amsmath,epsfig,slashed}
\usepackage{mathrsfs}
\usepackage{cellspace}

\begin{document}

\title{On the origin and geometry of anomaly-related poles \\ in Double Deeply Virtual Compton Scattering}

\author{Shohini Bhattacharya}
\email{shohinib@uconn.edu}
\affiliation{Department of Physics, University of Connecticut, Storrs, CT 06269, USA}

\author{Alexander Sang Yi}
\affiliation{Department of Physics, University of Connecticut, Storrs, CT 06269, USA}

\begin{abstract}
Anomaly-related poles---infrared poles whose residues carry
the gluonic operator structures associated with the chiral
or trace anomaly---have recently been reported in one-loop
box amplitudes for the gluon-initiated channel of double deeply
virtual Compton scattering (DDVCS). Their origin within the box amplitude
has remained unclear. We trace these poles to a subleading Landau singularity on the
boundary of the box's Feynman-parameter domain, where one parameter
vanishes, leaving an active triangle subdiagram.
We examine this singularity using dimensional regularization (DR),
nonzero momentum transfer (\(t\neq0\)), and a finite quark mass
(\(m\neq0\)). At \(t=m=0\), the boundary contribution has a power infrared
divergence in four dimensions. By analytic continuation, DR
conceals---but does not eliminate---this endpoint divergence,
which gives no \(1/\epsilon\) pole in the expansion about
\(d=4\). The same endpoint would diverge logarithmically at
the critical dimension \(d=6\), where the analytically
continued integral has a pole.
Nonzero \(t\) and finite \(m\) regulate the same region, yielding
the known \(1/t\) anomaly-related pole and a previously unreported
\(1/m^2\) behavior, respectively.
We also show that the pole residue depends on the
gluon-polarization projection. A transverse gluon-polarization projector can eliminate an
anomaly-related contribution that survives when the polarization
vectors are kept explicit. This may explain why it was absent
in earlier calculations. Finally, we present
complete one-loop results for polarized and unpolarized DDVCS
with all three infrared prescriptions. After summing the diagrams, we find that, in each of the
polarized and unpolarized amplitudes, the same evolution
kernel governs the coefficient of the infrared pole in DR and
those of the logarithms \(\ln(Q^2/(-t))\) and \(\ln(Q^2/m^2)\),
where \(Q^2\) denotes the hard photon-virtuality scale. This work establishes the common
geometric origin of these anomaly-related poles within one-loop box amplitudes.
\end{abstract}
\date{\today}

\maketitle

\section{Introduction}
Hard-exclusive Compton scattering provides a framework for studying the partonic structure of hadrons in quantum chromodynamics (QCD). In the generalized Bjorken regime, the amplitude separates into a perturbatively calculable hard coefficient and nonperturbative off-forward quark and gluon matrix elements, parameterized by generalized parton distributions (GPDs). GPDs correlate parton momentum with momentum transfer to the hadron; their moments give form factors, while at zero skewness their momentum-transfer dependence provides information about the transverse spatial distribution of partons\cite{Ji:1996ek,Radyushkin:1996nd,Collins:1998be,Ji:1998xh,Diehl:2003ny,Burkardt:2000za,Burkardt:2002hr}.

A growing direction in the GPD program is to use hard-exclusive
Compton amplitudes not only to map hadron structure, but also to
investigate how gluon fields shape some of QCD's most fundamental
phenomena. The chiral anomaly connects gluon topology to the
unexpectedly large mass of the \(\eta'\) through the pseudoscalar
operator \(F^{\mu\nu}\widetilde F_{\mu\nu}\)
\cite{Adler:1969gk,Bell:1969ts,Witten:1979vv,Veneziano:1979ec,
Tarasov:2020cwl,Tarasov:2021yll}. The trace anomaly connects the
scale dependence of QCD to hadron masses through the scalar gluonic
operator \(F^{\mu\nu}F_{\mu\nu}\), and is central to understanding
the origin of the proton mass
\cite{Ji:1994av,Hatta:2018sqd}. Hard-exclusive Compton amplitudes
therefore offer a way to access off-forward matrix elements of these
operators, bringing anomaly physics into the study of hadron
structure.

These gluonic operators also appear in amplitudes as residues of infrared poles. Such poles have long been studied in
three-point functions. In the chiral channel, the massless axial-vector/vector/vector
triangle produces a pole contribution to the off-forward matrix
element of the flavor-singlet axial current
\(J_5^\mu=\sum_f \bar q_f\gamma^\mu\gamma_5 q_f\),
where the sum runs over quark flavors. In
schematic form,
\begin{equation}
\langle p_2|J_5^\mu|p_1\rangle_{\mathrm{pole}}
\propto
\frac{\ell^\mu}{t}
\langle p_2|iF^{\alpha\beta}
\widetilde F_{\alpha\beta}|p_1\rangle;
\qquad
\ell=p_2-p_1, \qquad t = \ell^2.
\label{eq:intro-axial-anomaly-pole}
\end{equation}
Here, \(1/t\) is the infrared pole, whose residue carries the
gluonic operator structure of the chiral anomaly, a connection
discussed since the classic analyses of the axial triangle
\cite{Adler:1969gk,Bell:1969ts,Jaffe:1989jz}.

An analogous pole occurs in the trace-anomaly channel. In the
massless limit, the triangle contribution to the off-forward
energy-momentum-tensor $\Theta^{\mu\nu}$ matrix element has the schematic form
\begin{equation}
\langle p_2|\Theta^{\mu\nu}|p_1\rangle_{\mathrm{pole}}
\propto
\frac{1}{t}
\langle p_2|F^{\alpha\beta}F_{\alpha\beta}|p_1\rangle;
\qquad
\ell=p_2-p_1, \qquad t = \ell^2.
\label{eq:intro-trace-anomaly-pole}
\end{equation}
Giannotti and Mottola first identified the corresponding
trace-anomaly pole in QED~\cite{Giannotti:2008cv}.
Subsequent work studied its structure in other gauge theories \cite{Armillis:2009im,Armillis:2010qk}.
More recent dispersive analyses have examined how the pole strength
is shared between the pole and continuum as the kinematics and quark
mass vary \cite{Coriano:2025fom}.

At the level of scattering amplitudes, significant progress has
been made. Tarasov and Venugopalan traced the chiral-anomaly
contribution in polarized deep-inelastic scattering (DIS) to a
triangle mechanism within the one-loop box amplitude, in both
the Bjorken and Regge limits
\cite{Tarasov:2020cwl,Tarasov:2021yll}. In deeply virtual Compton
scattering (DVCS), Bhattacharya, Hatta, and Vogelsang identified
chiral-anomaly-related and trace-anomaly-related poles in the
polarized and unpolarized amplitudes, respectively, and provided
a prescription for relating their residues to the corresponding
gluonic operators
\cite{Bhattacharya:2022xxw,Bhattacharya:2023wvy}.
Going beyond perturbation theory, they discussed the implications
of these anomaly poles for the nonperturbative structure of
GPDs~\cite{Bhattacharya:2024geo}. Lending support to this
interpretation, a recent chiral quark soliton model calculation
finds that singlet pseudoscalar exchange gives a sizable
contribution to the flavor-singlet polarized GPD
\(\widetilde E\)~\cite{Minamiguchi:2026uaq,Minamiguchi:2026bbs}. Complementary operator-level
studies have examined gluon-state matrix elements, their relation
to GPDs, and the role of infrared regulators
\cite{Castelli:2024eza,Alvaro:2025gzl}.

In this work, we address the complementary question of why and how
the pole-producing region arises within the Compton loop integrals
and how its contribution depends on the infrared prescription.
We study this question in the
gluon channel of double deeply virtual Compton scattering (DDVCS)
using a systematic Landau analysis of the Passarino--Veltman
coefficient functions in the one-loop box amplitude.
The analysis identifies the singular region as the boundary of the box’s Feynman-parameter integration domain on which the parameter associated with the internal quark propagator joining the two photon vertices vanishes. In the Landau classification, this is a subleading
singularity: the propagator associated with the vanishing parameter
is inactive, while the remaining three propagators define a reduced
diagram with triangle topology. The original integral remains a
four-point function; the triangle describes the boundary
configuration that controls its singular behavior. Our Landau analysis thus identifies the triangle boundary responsible
for the pole in DDVCS. A related triangle mechanism was discussed
by Tarasov and Venugopalan in the worldline approach to polarized
DIS.
We further characterize this singular configuration and its
dependence on the infrared prescription in DDVCS.

We examine this boundary configuration using three infrared
prescriptions: dimensional regularization (DR), nonzero momentum
transfer (\(t\neq0\)), and a finite quark mass (\(m\neq0\)).
At \(t=m=0\), the relevant Passarino--Veltman coefficient
functions become singular as this boundary is approached.
Introducing a cutoff on the vanishing Feynman parameter reveals
a linear divergence in the inverse cutoff. DR assigns the
associated contribution zero by analytic continuation,
so this power divergence does not appear as a \(1/\epsilon\)
pole at \(d=4\). The analytically continued parameter integral
instead has a pole at \(d=6\), where the divergence becomes
logarithmic. This pole identifies the critical dimension of
the same boundary contribution.
Keeping \(t\) or \(m\) nonzero regulates this boundary region
and exposes contributions proportional to \(1/t\) or \(1/m^2\),
respectively. To our knowledge, the \(1/m^2\) behavior has not
previously been reported for these amplitudes. The triangle
boundary thus provides a common geometric origin for these
behaviors, and its role remains robust across all three
infrared prescriptions.

We also show that the pole residue depends on how the external
gluon polarizations are treated. Contracting the box amplitude
with a transverse projector can remove an anomaly-related
contribution that survives when the external gluon polarization
vectors are kept explicit. This sensitivity
may help explain the absence of such contributions in
earlier calculations. We also present complete one-loop results
for polarized and unpolarized DDVCS with all three infrared
prescriptions. After summing the diagrams, the corresponding terms
\(1/\epsilon_{\mathrm{IR}}\), \(\ln(Q^2/(-t))\), and
\(\ln(Q^2/m^2)\) multiply the same evolution kernel,
providing a consistency check across the three calculations.

The paper is organized as follows. Section~\ref{sec:preliminaries}
introduces the Passarino--Veltman notation and the Landau equations.
Section~\ref{sec:origin-anomaly-pole} identifies the triangle boundary
and examines how infrared regulators and gluon-polarization
projections affect its contribution.
Section~\ref{sec:complete-results} presents the complete one-loop
results for polarized and unpolarized DDVCS.
Section~\ref{sec:conclusions} summarizes our findings and outlines
prospects for future work.

\section{Preliminaries}
\label{sec:preliminaries}
\subsection{Passarino--Veltman functions}
\label{sec:pv-functions}
We express the one-loop tensor integrals in terms of the conventional
Passarino--Veltman coefficient functions \(A\), \(B\), \(C\), and \(D\);
see, for example,
Refs.~\cite{Passarino:1978jh,Ellis:2011cr}.
We work in
\begin{equation}
d=4-2\epsilon.
\end{equation}
We write \(\epsilon=\epsilon_{\mathrm{IR}}<0\) when regulating
infrared divergences and \(\epsilon=\epsilon_{\mathrm{UV}}>0\)
when regulating ultraviolet divergences.
In the Passarino--Veltman tensor decomposition, every pair of zero
indices represents one metric tensor, whereas the indices \(1,2,3\)
label the corresponding independent external momenta. We introduce the
nonnegative integers \(r,n_i=0,1,2,\ldots\), where \(r\) counts the
number of metric tensors and \(n_i\) counts the number of factors of
the \(i\)th external momentum. (Appendix~\ref{app:PV-notation} illustrates the index conventions
in detail using two-point functions as examples.) We then use the compact notation
\begin{align}
A_{0^{2r}}
&\equiv
A_{\underbrace{\scriptstyle 00\cdots0}_{2r}},
\nonumber\\
B_{0^{2r}1^{n_1}}
&\equiv
B_{\underbrace{\scriptstyle 00\cdots0}_{2r}
  \underbrace{\scriptstyle 11\cdots1}_{n_1}},
\nonumber\\
C_{0^{2r}1^{n_1}2^{n_2}}
&\equiv
C_{\underbrace{\scriptstyle 00\cdots0}_{2r}
  \underbrace{\scriptstyle 11\cdots1}_{n_1}
  \underbrace{\scriptstyle 22\cdots2}_{n_2}},
\nonumber\\
D_{0^{2r}1^{n_1}2^{n_2}3^{n_3}}
&\equiv
D_{\underbrace{\scriptstyle 00\cdots0}_{2r}
  \underbrace{\scriptstyle 11\cdots1}_{n_1}
  \underbrace{\scriptstyle 22\cdots2}_{n_2}
  \underbrace{\scriptstyle 33\cdots3}_{n_3}}.
\label{eq:pv-index-convention}
\end{align}
The corresponding tensor rank is
\begin{equation}
R=2r+\sum_i n_i.
\label{eq:PV-tensor-rank}
\end{equation}
In the Feynman-parameter representations below, the integers \(n_i\)
also determine the powers of the corresponding Feynman parameters.
The scalar functions are obtained by setting \(r=n_i=0\) and are
denoted by \(A_0\), \(B_0\), \(C_0\), and \(D_0\).

The general one-point coefficient is
\begin{equation}
A_{0^{2r}}(m_0)
=
\mu^{2\epsilon}e^{\gamma_E\epsilon}
\frac{(-1)^{r+1}}{2}
\Gamma(-1-r+\epsilon)
(m_0^2)^r
\left(
\frac{1}{m_0^2}
\right)^{-1+\epsilon}.
\label{eq:general-A}
\end{equation}
The symbol \(\mu\) denotes the dimensional regularization scale, and
\(\gamma_E\) is the Euler--Mascheroni constant. 
The general two-point coefficient is
\begin{align}
&B_{0^{2r}1^{n_1}}(s;m_0,m_1)
\nonumber\\
&\quad =
\mu^{2\epsilon}e^{\gamma_E\epsilon}
\frac{(-1)^{2+n_1+r}}{2^r}
\Gamma(\epsilon-r)
\int_0^1\mathrm{d}x\,
x^{n_1}
\bigl[F_B(x)-\mathrm{i}\varepsilon\bigr]^{-\epsilon+r},
\label{eq:general-B}
\end{align}
where
\begin{equation}
F_B(x)
=
s x^2
+\bigl(-s+m_1^2-m_0^2\bigr)x
+m_0^2.
\label{eq:FB}
\end{equation}
Here and below, \(s\), \(s_i\), and \(s_{ij}\) denote external
kinematic invariants. Their explicit expressions for the process
considered here will be specified in the following sections.
The general three-point coefficient is
\begin{align}
&C_{0^{2r}1^{n_1}2^{n_2}}
(s_1,s_{12},s_2;m_0,m_1,m_2)
\nonumber\\
&\quad =
\mu^{2\epsilon}e^{\gamma_E\epsilon}
\frac{(-1)^{3+n_1+n_2+r}}{2^r}
\Gamma(1+\epsilon-r)
\int_0^1\mathrm{d}x
\int_0^{1-x}\mathrm{d}y\,
x^{n_1}y^{n_2}
\bigl[F_C(x,y)-\mathrm{i}\varepsilon\bigr]^{-1-\epsilon+r},
\label{eq:general-C}
\end{align}
where
\begin{align}
F_C(x,y)
={}&
s_1x^2+s_2y^2+(s_1+s_2-s_{12})xy
+\bigl(-s_1+m_1^2-m_0^2\bigr)x
+\bigl(-s_2+m_2^2-m_0^2\bigr)y
+m_0^2.
\label{eq:FC}
\end{align}
The Feynman parameters \(x\) and \(y\) span the triangular domain
\begin{equation}
x\geq0,
\qquad
y\geq0,
\qquad
x+y\leq1.
\label{eq:C-domain}
\end{equation}
The general four-point coefficient is
\begin{align}
&D_{0^{2r}1^{n_1}2^{n_2}3^{n_3}}
(s_1,s_2,s_3,s_4;s_{12},s_{23};
m_0,m_1,m_2,m_3)
\nonumber\\
&\quad =
\mu^{2\epsilon}e^{\gamma_E\epsilon}
\frac{(-1)^{4+n_1+n_2+n_3+r}}{2^r}
\Gamma(2+\epsilon-r)
\int_0^1\mathrm{d}x
\int_0^{1-x}\mathrm{d}y
\int_0^{1-x-y}\mathrm{d}z\,
x^{n_1}y^{n_2}z^{n_3}
\bigl[F_D(x,y,z)-\mathrm{i}\varepsilon\bigr]^{-2-\epsilon+r},
\label{eq:general-D}
\end{align}
where
\begin{align}
F_D(x,y,z)
={}&
s_1x^2+s_{12}y^2+s_4z^2
+(s_1+s_{12}-s_2)xy
+(s_1-s_{23}+s_4)xz
+(s_{12}-s_3+s_4)yz
\nonumber\\
&+\bigl(-s_1+m_1^2-m_0^2\bigr)x
+\bigl(-s_{12}+m_2^2-m_0^2\bigr)y
+\bigl(-s_4+m_3^2-m_0^2\bigr)z
+m_0^2.
\label{eq:FD}
\end{align}
The Feynman parameters \(x,y,z\) span the tetrahedral domain
\begin{equation}
x\geq0,
\qquad
y\geq0,
\qquad
z\geq0,
\qquad
x+y+z\leq1.
\label{eq:D-domain}
\end{equation}

For the present analysis, the relation between the three-point and
four-point functions is particularly important. On the boundary
\(z=0\) of the box parameter domain,
\begin{align}
F_D(x,y,0)
={}&
s_1x^2+s_{12}y^2+(s_1+s_{12}-s_2)xy
+\bigl(-s_1+m_1^2-m_0^2\bigr)x
+\bigl(-s_{12}+m_2^2-m_0^2\bigr)y
+m_0^2.
\label{eq:FD-z-zero}
\end{align}
This is precisely the three-point (Symanzik) polynomial in
Eq.~\eqref{eq:FC}, with the triangle invariants identified as
\(s_1^{(C)}=s_1\), \(s_{12}^{(C)}=s_2\), and
\(s_2^{(C)}=s_{12}\), while the masses \(m_0,m_1,m_2\)
remain unchanged. Thus,
\begin{equation}
F_D(x,y,0)
=
\left.
F_C(x,y)
\right|_{
(s_1^{(C)},s_{12}^{(C)},s_2^{(C)})
=
(s_1,s_2,s_{12})
}.
\label{eq:FD-to-FC}
\end{equation}
In the argument ordering of Eq.~\eqref{eq:general-C}, this boundary
is therefore associated with the triangle coefficient
\begin{equation}
C_{0^{2r}1^{n_1}2^{n_2}}
\bigl(s_1,s_2,s_{12};m_0,m_1,m_2\bigr).
\label{eq:embedded-triangle}
\end{equation}
Thus, the \(z=0\) boundary of the box is described by an embedded
triangle subdiagram. This relation will be important for the
anomaly-related pole studied below.

\subsection{Landau equations}
\label{sec:landau-preliminaries}
Feynman integrals are generally complicated functions of the external
kinematic invariants, and their singularities can be difficult to
identify by direct integration. The Landau equations provide necessary
conditions for locating possible singularities directly from the
propagator denominators~\cite{Landau:1959fi}. They may be expressed
either in momentum space or, equivalently, in Feynman-parameter space.

Consider a one-loop integral with \(N\) propagators, labeled by
\(i=0,\ldots,N-1\), with denominators 
\begin{equation}
\mathscr D_i(k)+\mathrm{i}\varepsilon,
\qquad
\mathscr D_i(k)=(k+\kappa_i)^2-m_i^2,
\qquad
\varepsilon\to0^+,
\label{eq:general-propagators}
\end{equation}
where \(k\) is the loop momentum and the \(\kappa_i\) are fixed
combinations of external momenta. Introducing nonnegative Feynman parameters \(\alpha_i\), the
momentum-space Landau equations are
\begin{align}
\alpha_i\mathscr D_i(k)&=0,
\qquad i=0,\ldots,N-1,
\label{eq:landau-onshell}\\
\sum_{i=0}^{N-1}
\alpha_i(k+\kappa_i)^\mu&=0.
\label{eq:landau-stationarity}
\end{align}
If \(\alpha_i>0\), the corresponding propagator is active and must
satisfy \(\mathscr D_i(k)=0\). If \(\alpha_i=0\), that propagator is
inactive. The second equation is the
stationarity condition with respect to the loop momentum and follows
from
\(\partial_{k_\mu}\sum_i\alpha_i\mathscr D_i(k)=0\).

With the normalization
\begin{equation}
\alpha_i\geq0,
\qquad
\sum_{i=0}^{N-1}\alpha_i=1,
\label{eq:feynman-simplex}
\end{equation}
the Feynman parameters span an \((N-1)\)-dimensional simplex.
After eliminating \(\alpha_0\), the Landau equations at an
interior point take the equivalent parameter-space form
\begin{equation}
\mathcal F=0,
\qquad
\frac{\partial\mathcal F}{\partial\alpha_j}=0,
\qquad
j=1,\ldots,N-1,
\label{eq:parametric-landau}
\end{equation}
where the derivatives are taken with respect to the remaining
independent parameters. Note that \(\mathcal F\) is the usual
one-loop Symanzik polynomial,
\begin{equation}
\mathcal F(\alpha)
=
\sum_{i=0}^{N-1}\alpha_i m_i^2
-
\sum_{0\leq i<j\leq N-1}
\alpha_i\alpha_j(\kappa_i-\kappa_j)^2,
\label{eq:F-polynomial}
\end{equation}
whose explicit two-, three-, and four-point forms are the
polynomials \(F_B\), \(F_C\), and \(F_D\) introduced above.
If one or more parameters vanish, the Landau conditions apply
to \(\mathcal F\) restricted to the corresponding boundary face,
with stationarity imposed only along that face. Such solutions
are governed by lower-point subdiagrams. In particular, the
\(z=0\) boundary of the box is described by an embedded triangle,
as shown in Eq.~\eqref{eq:FD-to-FC}.

\section{Origin and geometry of the anomaly-related pole}
\label{sec:origin-anomaly-pole}
\begin{figure}[t]
\centering
\includegraphics[width=0.7\linewidth]{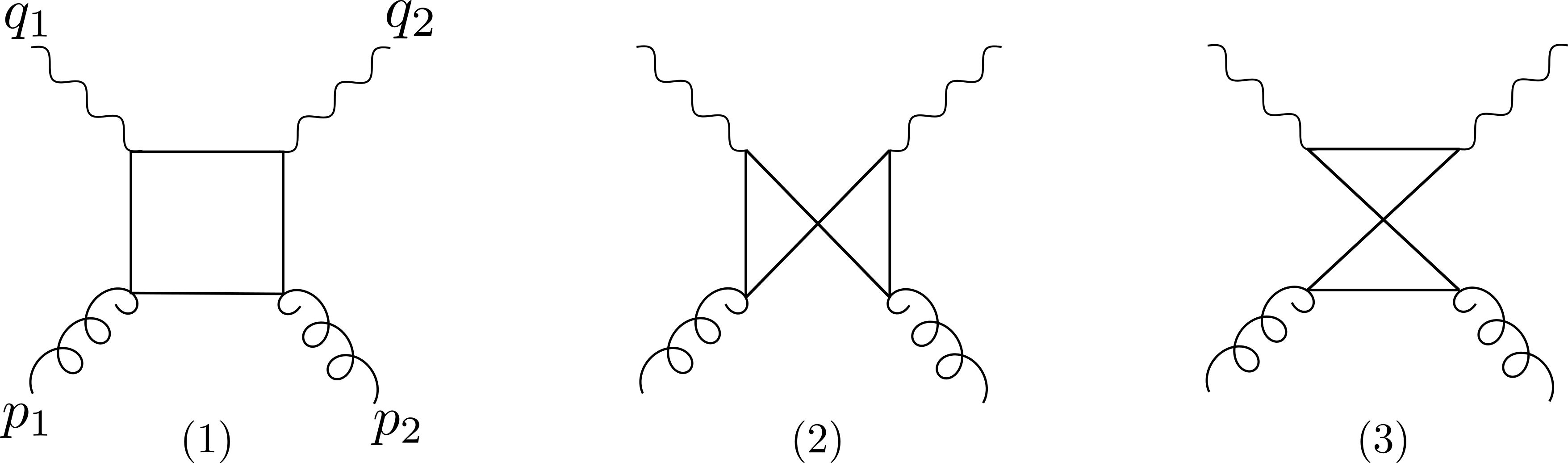}
\caption{The three gluon-channel diagrams relevant for DDVCS: the box (1), cat's-eye
(2), and hourglass (3) topologies.}

\label{fig:three-gluon-diagrams}
\end{figure}

The primary goal of this section is to explain how the \(1/t\)
poles reported in the gluon channel of polarized and unpolarized
DDVCS arise from the one-loop
integrals~\cite{Bhattacharya:2022xxw,Bhattacharya:2023wvy}.
Throughout this section (and the remaining sections), we use ``polarized'' and ``unpolarized''
interchangeably with ``antisymmetric'' and ``symmetric,''
respectively, referring to the symmetry of the Compton amplitude
under interchange of the photon Lorentz indices.
The residues of these poles in the polarized and unpolarized
amplitudes carry the gluonic operator structures associated with
the chiral and trace anomalies, respectively. Despite these
distinct residue structures, both poles arise from the same
subleading Landau singularity of the one-loop box. At this
singularity, the Feynman parameter associated with the quark
propagator joining the two photon vertices vanishes. This places
the singularity on a boundary face of the box's Feynman-parameter
domain, where the other three propagators form a triangle
subdiagram. We examine this boundary region using three infrared
prescriptions: dimensional regularization (DR), nonzero
momentum transfer ($t \neq 0$), and a finite quark mass ($m \neq 0$).
The comparison shows how each prescription treats the same
singular region.

Below, we carry out the calculation for the \(s\)-channel box,
shown as diagram~(1) in Fig.~\ref{fig:three-gluon-diagrams}.
The same calculation and conclusions apply to the \(u\)-channel
hourglass, shown as diagram~(3) in the same figure. By contrast, the cat's-eye topology shown as diagram~(2)
does not generate an anomaly-related pole.
To expose the dependence on the momentum transfer, we first state the
partonic kinematics at finite \(t\). We define the average photon and
parton momenta as (see Refs.~\cite{Bhattacharya:2022xxw,Bhattacharya:2023wvy})
\begin{equation}
q=\frac{q_1+q_2}{2},
\qquad
p=\frac{p_1+p_2}{2},
\label{eq:average-partonic-momenta}
\end{equation}
and parametrize the incoming and outgoing momenta as
\begin{equation}
q_1=q+\frac{\ell}{2},
\qquad
q_2=q-\frac{\ell}{2},
\qquad
p_1=p-\frac{\ell}{2},
\qquad
p_2=p+\frac{\ell}{2}.
\label{eq:partonic-momenta}
\end{equation}
The momentum transfer is therefore
\begin{equation}
\ell=q_1-q_2=p_2-p_1,
\qquad
t=\ell^2.
\label{eq:partonic-momentum-transfer}
\end{equation}
The incoming and outgoing partons are taken to be massless,
\(p_1^2=p_2^2=0\), which implies
\begin{equation}
p^2=-\frac{t}{4},
\qquad
p\cdot\ell=0.
\label{eq:partonic-onshell-conditions}
\end{equation}
Following the definitions used at the hadronic level in
Ref.~\cite{Bhattacharya:2023wvy}, we introduce the corresponding
partonic variables
\begin{equation}
Q^2=-q^2,
\qquad
\hat{x}
=
\frac{Q^2}{2p\cdot q},
\qquad
\hat{\xi}
=
\frac{q_2^2-q_1^2}{4p\cdot q}.
\label{eq:partonic-variables}
\end{equation}
The photon virtualities can then be written as
\begin{equation}
q_1^2 \equiv - Q_a^2
=
-Q^2\frac{\hat{x}+\hat{\xi}}{\hat{x}}
+\frac{t}{4},
\qquad
q_2^2 \equiv - Q_b^2
=
-Q^2\frac{\hat{x}-\hat{\xi}}{\hat{x}}
+\frac{t}{4}.
\label{eq:photon-virtualities}
\end{equation}
For DDVCS,
\(\hat{x}\) and \(\hat{\xi}\) are treated as independent
partonic variables. The \(t=0\) kinematics used in DR
and finite-quark-mass (\(m\neq0\)) calculations follow directly from the above expressions.

With the kinematics specified, we now turn to the partonic
Compton amplitude and consider two treatments of the external
gluon polarizations. First, we retain the polarization vectors
explicitly and write
\begin{equation}
\mathcal T^{\mu\nu}
=
{\mathcal T}^{\mu\nu\alpha\beta}\,
\epsilon_\alpha(p_1)\epsilon_\beta^*(p_2).
\label{eq:Compton-gluon-polarizations}
\end{equation}
Here, \(\mu\) and \(\nu\) are the Lorentz indices of the incoming
and outgoing photons in Fig.~\ref{fig:three-gluon-diagrams},
respectively. The vectors \(\epsilon_\alpha(p_1)\) and
\(\epsilon_\beta^*(p_2)\) describe the polarizations of the
incoming and outgoing gluons, respectively.
Second, we contract the gluon Lorentz indices \(\alpha\) and
\(\beta\) of the amplitude with the transverse projector
\(g_{\perp\,\alpha\beta}\) (see Eq.~(\ref{eq:transverse-projector})), giving
\begin{equation}
\mathcal T_\perp^{\mu\nu}
=
{\mathcal T}^{\mu\nu\alpha\beta}\,g_{\perp\,\alpha\beta}.
\label{eq:Compton-transverse-projection}
\end{equation}
Below, we identify the anomaly-related terms with the
gluon polarization vectors kept explicit and examine how the
transverse projection affects their residues. Finally, we note that the diagrams are evaluated using the
Mathematica package \textsc{Package-X}~\cite{Patel:2015tea}.

\subsection{Dimensional Regularization (DR)}
\label{subsec:dimensional-regularization}
\subsubsection{Identifying the triangle singularity in the box}
We begin with the \(t=m=0\) calculation in DR. The relevant loop integrals have been
evaluated previously~\cite{Ji:1997nk,Belitsky:1997rh,Pire:2011st}.
Here we examine the analytic structure of their
Passarino--Veltman coefficient functions to locate the singular
region within the box. In the Feynman-parameter representation,
this region lies near the boundary \(z=0\). As we will show, in
four dimensions a lower cutoff \(\delta\) on \(z\) exposes a
power divergence proportional to \(1/\delta\). Analytic
continuation in DR conceals this endpoint divergence, which
produces no \(1/\epsilon\) pole in the expansion about \(d=4\).
In the following subsections, we show that nonzero momentum
transfer and a finite quark mass regulate the same boundary
region, yielding the anomaly-related \(1/t\) pole and the
\(1/m^2\) behavior, respectively.

After tensor reduction, both the symmetric and antisymmetric parts of
the Compton amplitude, \(\mathcal T_S^{\mu\nu}\) and
\(\mathcal T_A^{\mu\nu}\), contain four-point coefficient functions of
the form
\begin{equation}
D_I
\left(
0,0,-Q_b^2,-Q_a^2;
0,s;
0,0,0,0
\right),
\qquad
I\in\mathcal I_D,
\label{eq:antisymmetric-D-family}
\end{equation}
where \(I\) is the Passarino--Veltman coefficient multi-index and
\begin{align}
\mathcal I_D
=\Bigl\{&
0,\,
1,\,
2,\,
3,\,
11,\,
12,\,
13,\,
22,\,
23,\,
33,
\nonumber\\
&
112,\,
113,\,
122,\,
123,\,
133,\,
223,\,
233,\,
333,
\nonumber\\
&
00,\,
001,\,
002,\,
003
\Bigr\}.
\label{eq:D-index-set}
\end{align}
Although these coefficients differ in tensor rank and numerator
structure, they share the same four-point denominator polynomial (see below) and
therefore the same candidate Landau singularities. Their tensor
indices determine the strength with which a singular region
contributes, but not its location.

\begin{figure}
\centering
\includegraphics[width=0.75\linewidth]{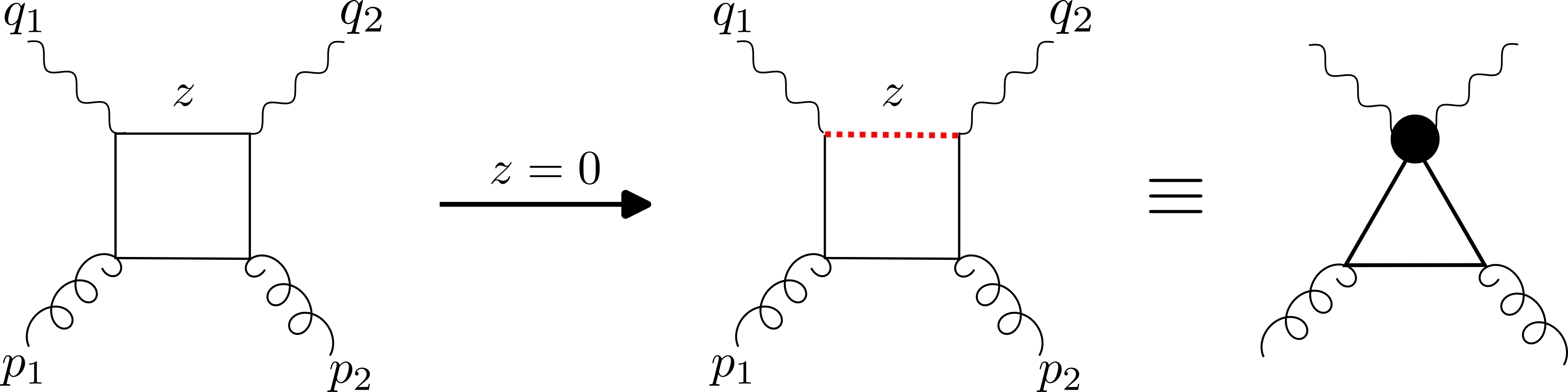}
\caption{Box-to-triangle reduction on the Landau boundary \(z=0\).
The red dashed line marks the quark propagator joining the two
photon vertices, whose Feynman parameter \(z\) vanishes on this
boundary. The remaining three propagators then form the triangle
subdiagram responsible for the anomaly-related Landau singularity.
The same boundary reduction applies to the \(u\)-channel
hourglass diagram.}
\label{fig:z-zero-reduction}
\end{figure}
At \(t=0\), the four-point coefficient functions \(D_I\) share the
denominator polynomial
\begin{equation}
F_D(x,y,z)
=
z\bigl[a(x,y)-Q_a^2z\bigr],
\label{eq:FD-t-zero}
\end{equation}
where
\begin{equation}
a(x,y)
=
Q_a^2
-\bigl(Q_a^2+s\bigr)x
+\bigl(Q_b^2-Q_a^2\bigr)y.
\label{eq:a-xy}
\end{equation}
For the momentum routing used here, \(z\) is the Feynman parameter
associated with the internal quark propagator joining the two photon
vertices. The overall factor of \(z\) in \(F_D\) implies that
\begin{equation}
F_D(x,y,0)=0,
\qquad
\left.\frac{\partial F_D}{\partial x}\right|_{z=0}
=
\left.\frac{\partial F_D}{\partial y}\right|_{z=0}
=0
\label{eq:z-zero-landau}
\end{equation}
throughout the face \(z=0\). These are the Landau conditions
on that face, with stationarity imposed in the \(x\) and \(y\)
directions. The derivative in the \(z\) direction is
\begin{equation}
\left.\frac{\partial F_D}{\partial z}\right|_{z=0}
=
a(x,y).
\label{eq:normal-FD-derivative}
\end{equation}
Since \(z\) is fixed to zero on the face, this derivative is
unconstrained by the boundary Landau conditions.
In the Landau classification, the solution at \(z=0\) is a
subleading singularity of the box. The propagator associated
with the vanishing parameter \(z\) is inactive, while the
remaining three propagators form the triangle subdiagram shown
in Fig.~\ref{fig:z-zero-reduction}. The loop integral is still
a four-point function. The triangle identifies the boundary
configuration that controls its singular behavior and, as we
show below, gives rise to the anomaly-related \(1/t\) and $1/m^2$ pole in
DDVCS. A related triangle mechanism was discussed by Tarasov
and Venugopalan in their worldline analysis of polarized
DIS~\cite{Tarasov:2020cwl,Tarasov:2021yll}.

Mathematically, the box-to-triangle reduction follows by
restricting the box denominator polynomial to the Landau
boundary \(z=0\). As shown in Eq.~\eqref{eq:FD-to-FC},
setting \(z=0\) removes the terms containing the Feynman
parameter of the inactive propagator and leaves the triangle
polynomial of the three active propagators,
\begin{equation}
    F_D(x,y,0)=F_C(x,y) .
\end{equation}
For the massless, zero-transfer kinematics of the box in
Eq.~\eqref{eq:antisymmetric-D-family}, all three external
invariants and all three internal masses of the triangle vanish.
The associated scalar three-point function is $C_0(0,0,0;0,0,0)$.
The corresponding tensor three-point coefficients have the same
kinematic arguments.
By infrared power counting, the scalar triangle is power
divergent in four dimensions. The calculation below shows
how this infrared behavior is explicitly realized in the original box
integral as \(z\to0\). We examine
the four-point integrand as \(z\to0\) and find that a lower
cutoff \(\delta\) on \(z\) exposes a \(1/\delta\) divergence
in the scalar box. Analytic continuation in DR conceals this
power divergence, which produces no \(1/\epsilon\) pole in
the expansion about \(d=4\). At the critical dimension \(d=6\),
the same endpoint diverges logarithmically, giving a pole in
the analytically continued endpoint integral.

We now examine how the Feynman-parameter integrands of the
four-point coefficient functions \(D_I\) behave as \(z\to0\). In
\(d=4-2\epsilon\), a general coefficient has the form
\begin{align}
&D_{0^{2r}1^{n_1}2^{n_2}3^{n_3}}
\propto
\int\mathrm{d}x\,\mathrm{d}y\,\mathrm{d}z\,
x^{n_1}y^{n_2}z^{n_3}
\bigl[F_D(x,y,z)-\mathrm{i}\varepsilon
\bigr]^{-2-\epsilon+r}.
\label{eq:D-boundary-representation}
\end{align}
At a generic point of the \(z=0\) boundary, where \(a(x,y)\neq0\),
Eq.~\eqref{eq:FD-t-zero} gives
\begin{equation}
F_D(x,y,z)
=
a(x,y)z+\mathcal O(z^2).
\label{eq:FD-generic-boundary}
\end{equation}
Thus, for power counting normal to the boundary, the integrand behaves
as
\begin{equation}
z^{n_3}
\bigl[F_D-\mathrm{i}\varepsilon\bigr]^{-2-\epsilon+r}
\sim
\bigl[a(x,y)\bigr]^{-2-\epsilon+r}
z^{\,n_3-2-\epsilon+r}.
\label{eq:D-boundary-scaling}
\end{equation}
Here the causal prescription is understood; the displayed scaling
describes the endpoint power away from the zero of \(a(x,y)\).

For the scalar coefficient, \(r=n_3=0\), the generic \(z=0\)
boundary gives the endpoint behavior
\begin{equation}
\int_0 \mathrm{d}z\,z^{-2-\epsilon}.
\label{eq:linear-z-divergence}
\end{equation}
At \(d=4\), or \(\epsilon=0\), this endpoint is not integrable in the
ordinary sense. Introducing a dimensionless lower cutoff \(\delta\)
on the Feynman parameter makes the degree of divergence explicit:
\begin{equation}
\int_\delta^1\frac{\mathrm{d}z}{z^2}
=
\frac{1}{\delta}-1.
\label{eq:linear-z-cutoff}
\end{equation}
Thus, in a cutoff description of the parameter endpoint, this
contribution diverges linearly in \(1/\delta\).

In DR, on the other hand, we evaluate the model endpoint
integral in its region of convergence, \(\operatorname{Re}\epsilon<-1\),
where
\begin{equation}
\int_0^1\mathrm{d}z\,z^{-2-\epsilon}
=
-\frac{1}{1+\epsilon}.
\label{eq:linear-z-convergent-region}
\end{equation}
Analytic continuation to \(\epsilon=0\) gives a finite value,
even though the original four-dimensional endpoint integral
diverges as \(1/\delta\) when the lower cutoff \(\delta\) tends
to zero. DR thus conceals the
four-dimensional power divergence, which gives no
\(1/\epsilon\) pole in the expansion about \(d=4\).
The continued endpoint integral has a pole at \(\epsilon=-1\),
which singles out \(d=6\) as the critical dimension for this
endpoint. At \(d=6\), the integrand scales as \(1/z\), and the
endpoint integral diverges logarithmically. The pole at \(d=6\)
therefore marks the transition from power-divergent behavior
below six dimensions to an integrable endpoint above six
dimensions.
As we show below, nonzero momentum transfer and a finite quark
mass regulate the same boundary region, yielding \(1/t\) and
\(1/m^2\) terms, respectively. After summing the relevant diagrams
and projecting onto the appropriate tensor structures, the
residues of these terms carry the gluonic operator structures
associated with \(F^{\mu\nu}F_{\mu\nu}\) and
\(F^{\mu\nu}\widetilde{F}_{\mu\nu}\). Note that at a generic point on the
boundary, additional powers of \(z\) in the numerator or metric
contributions with \(r>0\) soften the endpoint behavior, as shown
in Eq.~\eqref{eq:D-boundary-scaling}.

\subsubsection{Effect of the transverse gluon projector}
The calculation of the unpolarized DVCS amplitude
in Ref.~\cite{Pire:2011st} uses the following transverse
polarization sum for the external gluons:
\begin{align}
g_\perp^{\alpha\beta}
&=
g^{\alpha\beta}
-\frac{1}{Q^2}
\left(q^\alpha+\frac{Q^2}{p\cdot q}p^\alpha\right)
\left(q^\beta+\frac{Q^2}{p\cdot q}p^\beta\right)
+\frac{q^\alpha q^\beta}{Q^2}.
\label{eq:transverse-projector}
\end{align}
We loosely refer to the amplitude obtained by contracting the two
gluon Lorentz indices with \(g_\perp^{\alpha\beta}\) in
Eq.~\eqref{eq:Compton-transverse-projection} as the ``projected
amplitude''. We compare it with the amplitude in which the
polarization product \(\epsilon_\alpha(p_1)\epsilon_\beta^*(p_2)\)
is kept explicit, as in Eq.~\eqref{eq:Compton-gluon-polarizations}.
As we show below, contracting with \(g_\perp^{\alpha\beta}\)
changes the numerator and can modify the residue associated with
the Landau singularity identified by the propagator denominators.

To illustrate this effect, consider the box diagram in
Fig.~\ref{fig:three-gluon-diagrams}. We use the loop momentum
routing of Ref.~\cite{Pire:2011st}: the left and right vertical
lines carry \(k+\xi p\) and \(k-\xi p\), respectively, and
the upper and lower horizontal lines carry \(k+q\) and \(k-p\),
respectively. This routing follows from the lightlike momentum convention of
Ref.~\cite{Pire:2011st},
\begin{equation}
\ell^\mu\longrightarrow -2\xi p^\mu,
\end{equation}
so \(\ell^2=0\), and
\begin{equation}
k+\xi p=k-\frac{\ell}{2},
\qquad
k-\xi p=k+\frac{\ell}{2}.
\end{equation}
Following the notation of Ref.~\cite{Pire:2011st}, we denote the
left and right vertical propagator denominators by \(D_-\) and
\(D_+\), respectively, and the upper horizontal denominator by
\(D_q\). The lower horizontal propagator denominator is
\(D_p\).
Thus, the four denominators read
\begin{equation}
D_\pm=\left(k\pm\frac{\ell}{2}\right)^2,
\qquad
D_q=(k+q)^2,
\qquad
D_p=(k-p)^2.
\label{eq:transverse-projector-propagators}
\end{equation}

The
transverse projection gives a loop numerator of the form
\begin{equation}
N=(k\cdot q)\bigl[A k^2+B(k\cdot p)\bigr],
\label{eq:transverse-projected-numerator}
\end{equation}
where \(A\) and \(B\) are kinematic coefficients. Thus the projected
box contribution is the loop integral
\begin{equation}
\mathcal I_T
\equiv
\int_k
\frac{(k\cdot q)\bigl[A k^2+B(k\cdot p)\bigr]}
     {D_+D_-D_qD_p}
=\mathcal I_{T1}+\mathcal I_{T2},
\label{eq:transverse-projected-loop}
\end{equation}
where
\begin{equation}
\mathcal I_{T1}
\equiv
A\int_k\frac{(k\cdot q)k^2}{D_+D_-D_qD_p},
\qquad
\mathcal I_{T2}
\equiv
B\int_k\frac{(k\cdot q)(k\cdot p)}{D_+D_-D_qD_p}.
\label{eq:transverse-projected-loop-pieces}
\end{equation}
Here \(\int_k\) denotes the dimensionally regulated loop integration;
the causal \(+\mathrm{i} \varepsilon\) prescription in each propagator is
understood.

\begin{figure}
\centering
\includegraphics[width=0.7\linewidth]{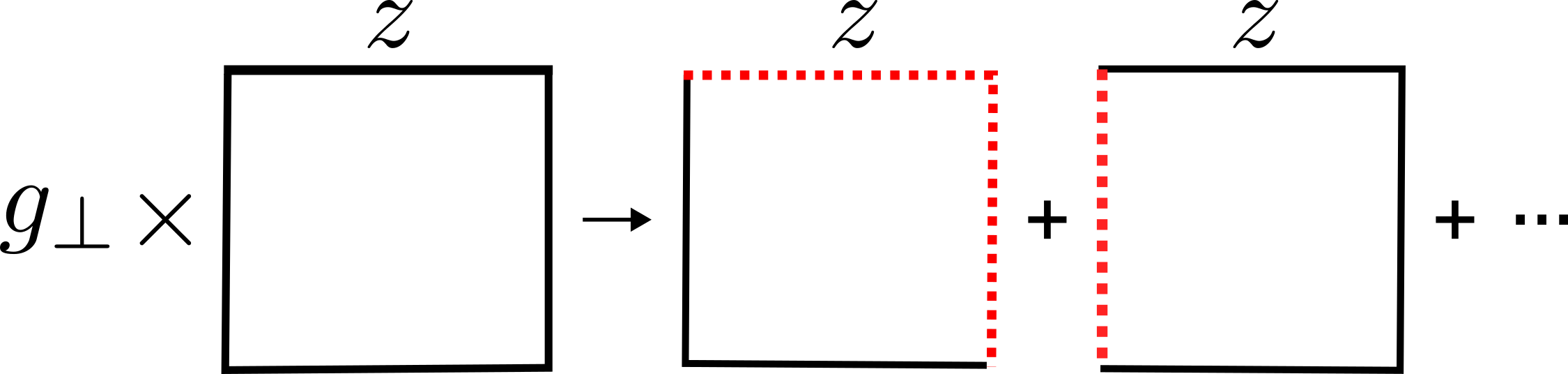}
  \caption{Schematic illustration of the effect of a transverse
gluon-polarization projector (\(g_\perp^{\alpha\beta}\) or
\(\epsilon_\perp^{\alpha\beta}\)) on the box numerator.
In the projected integrand, numerator factors cancel
propagator denominators, allowing the box contribution to be
expressed as a sum of bubble and triangle integrals. The
resulting combination is free of the anomaly-related pole.}
  \label{fig:proj}
\end{figure}

The scalar products in the numerator can be rewritten in terms of the
denominators:
\begin{align}
k\cdot q
&=
\frac{1}{2}D_q-\frac{1}{4}D_+-\frac{1}{4}D_-
-\frac{q^2}{2},
\nonumber\\
k^2
&=\frac{1}{2}(D_++D_-),
\nonumber\\
k\cdot p
&=\frac{1}{4}(D_++D_-)-\frac{1}{2}D_p .
\label{eq:transverse-projector-numerator-identities}
\end{align}
Inserting these identities into the two terms in
Eq.~\eqref{eq:transverse-projected-loop-pieces} yields
\begin{align}
\mathcal I_{T1}=A\int_k\Bigg[
&\frac{1}{4D_-D_p}
+\frac{1}{4D_+D_p}
-\frac{1}{4D_qD_p}
\nonumber\\
&-\frac{D_+}{8D_-D_qD_p}
-\frac{D_-}{8D_+D_qD_p}
-\frac{q^2}{4D_-D_qD_p}
-\frac{q^2}{4D_+D_qD_p}
\Bigg],
\label{eq:transverse-projector-I1-reduction}
\\[2mm]
\mathcal I_{T2}=B\int_k\Bigg[
&\frac{1}{8D_-D_p}
+\frac{1}{8D_+D_p}
-\frac{1}{4D_+D_-}
-\frac{1}{8D_qD_p}
+\frac{1}{8D_-D_q}
+\frac{1}{8D_+D_q}
\nonumber\\
&-\frac{D_+}{16D_-D_qD_p}
-\frac{D_-}{16D_+D_qD_p}
-\frac{q^2}{8D_-D_qD_p}
-\frac{q^2}{8D_+D_qD_p}
+\frac{q^2}{4D_+D_-D_q}
\Bigg].
\label{eq:transverse-projector-I2-reduction}
\end{align}
The numerator factors cancel one or more propagators, reducing
the projected loop integral to a linear combination of bubble (first line of the above equations)
and triangle (second line of the above equations) integrals. The triangle
integrals in this reduction do not generate the power infrared
divergence associated with the \(z=0\) boundary of the original
box, where the quark propagator joining the two photon vertices
becomes inactive. See Fig.~\ref{fig:proj} for a schematic illustration of this. So, in essence, the \(g_\perp^{\alpha\beta}\) projection
changes the numerator and hence the residue with which this boundary
contributes; a suppressed or vanishing residue in the projected
amplitude does not mean that the triangle boundary is absent. The
bubble and triangle integrals in
Eqs.~\eqref{eq:transverse-projector-I1-reduction} and
\eqref{eq:transverse-projector-I2-reduction} must accordingly be
understood as the projected combination, rather than as evidence that
the underlying Landau surface has disappeared.

Finally, we find the same pattern when we apply the analogous
transverse projection to the polarized DVCS amplitude. In this
case, the transverse projector is
\begin{equation}
\epsilon_{\perp}^{\alpha\beta}
\propto
\epsilon^{\alpha\beta\rho\sigma}p_{\rho}q_{\sigma}.
\end{equation}
We refrain from showing here the explicit reduction of the
projected loop integrals to bubble and triangle integrals, as
it follows the same pattern as in the unpolarized case. 

The calculations in this section establish two results that, to our
knowledge, have not previously been made explicit. First, we identify the \(z=0\) boundary of the box's
Feynman-parameter domain as the source of a power infrared
divergence in the relevant Passarino--Veltman coefficients. A lower cutoff on \(z\) makes this
divergence explicit in four dimensions. In DR, analytic continuation conceals the power
divergence, which produces no \(1/\epsilon\) pole in the expansion
about \(d=4\). 
Second, we show that contracting the box amplitude with a transverse
gluon projector can suppress the associated pole residue through its
effect on the numerator. The propagator denominators still satisfy
the Landau conditions on the same boundary. This sensitivity to the
polarization projection may help explain why the contribution was
absent in earlier calculations.
In the following subsections, we show that nonzero momentum transfer
and a finite quark mass regulate this same boundary region, yielding
the anomaly-related \(1/t\) pole and the \(1/m^2\) behavior,
respectively. The triangle mechanism is thus robust across all three
infrared prescriptions: the same boundary controls the singular
behavior in each case.

\subsection{Nonzero momentum transfer (\(t\neq0)\)}
We first regulate the \(z=0\) boundary by keeping the momentum transfer
nonzero. The tensor-reduced amplitude contains the four-point
coefficient functions
\begin{equation}
D_I\left(
0,0,-Q_b^2,-Q_a^2;
t,s;
0,0,0,0
\right),
\qquad I\in\mathcal I_D,
\label{eq:nonzero-t-D-family}
\end{equation}
with the coefficient-index set \(\mathcal I_D\) defined in
Eq.~\eqref{eq:D-index-set}. For this family, the denominator
polynomial is
\begin{equation}
F_D(x,y,z)
=
a(x,y)z+b(x,y)t+cz^2+d(x,y)tz,
\label{eq:FD-small-t-decomposition}
\end{equation}
where \(a(x,y)\) is defined in Eq.~\eqref{eq:a-xy}, and
\begin{equation}
b(x,y)=-y(1-x-y),
\qquad
c=-Q_a^2,
\qquad
d(x,y)=y.
\label{eq:small-t-bcd}
\end{equation}
At \(t=0\), this reduces to
\begin{equation}
F_D(x,y,z)=z\bigl[a(x,y)-Q_a^2z\bigr],
\end{equation}
which vanishes along the full \(z=0\) face. This is the boundary
associated with the reduced triangle shown in Fig.~\ref{fig:z-zero-reduction}.

To analyze the contribution from this boundary, choose a fixed,
small \(z_0\), independent of \(t\), and restrict \(z\) to
\(0\leq z\leq z_0\). The value \(z_0\) only marks the upper end of the
local boundary region; it is not a physical scale or a lower cutoff.
At generic fixed \(x,y\), with \(a(x,y),b(x,y)\neq0\)\footnote{These estimates apply at generic fixed \(x,y\). In particular, the
fixed-\(a\) expansion is not uniform near \(a(x,y)=0\); that region
must be analyzed separately in the full parameter integral. The
resulting strip estimate describes another scaling region of the same
integral and should not be added to the fixed-\(a\) estimate as an
independent contribution.}, the terms
\(a(x,y)z\) and \(b(x,y)t\) become comparable at
\begin{equation}
z\sim z_t,
\qquad
z_t\equiv
\left|\frac{b(x,y)t}{a(x,y)}\right|.
\label{eq:zt-crossover}
\end{equation}
The scale \(z_t\) divides the boundary region into two ranges in which
different terms dominate. The original integral continues to run from
\(z=0\); we split it at \(z_t\) only to estimate the contribution on
each side of the crossover.

For \(r=0\), the scalar coefficient has a local integrand
proportional to \(F_D^{-2}\). Keeping the two competing terms gives
\begin{align}
I_t
&\equiv
\int_0^{z_0}
\frac{\mathrm{d}z}
     {\bigl[a(x,y)z+b(x,y)t\bigr]^2}
\nonumber\\
&=
\frac{1}{a(x,y)b(x,y)t}
-\frac{1}{
a(x,y)\bigl[a(x,y)z_0+b(x,y)t\bigr]}.
\label{eq:nonzero-t-leading-pole}
\end{align}
For generic \(a(x,y)b(x,y)\neq0\), the second term remains finite as
\(t\to0\), whereas the first is proportional to \(1/t\).
The two sides of the crossover both contribute at this inverse-power
order. For \(0\leq z\lesssim z_t\), the term \(b(x,y)t\) dominates, so
\begin{equation}
\int_0^{z_t}
\frac{\mathrm{d}z}
     {\bigl[a(x,y)z+b(x,y)t\bigr]^2}
\sim
\frac{z_t}{[b(x,y)t]^2}
\sim \frac{1}{|a(x,y)b(x,y)|\,|t|}.
\label{eq:t-below-crossover}
\end{equation}
For \(z_t\lesssim z\leq z_0\), the term \(a(x,y)z\) dominates, giving
\begin{equation}
\int_{z_t}^{z_0}
\frac{\mathrm{d}z}
     {\bigl[a(x,y)z+b(x,y)t\bigr]^2}
\sim
\frac{1}{a(x,y)^2}
\left(\frac{1}{z_t}-\frac{1}{z_0}\right)
\sim \frac{1}{|a(x,y)b(x,y)|\,|t|}.
\label{eq:t-above-crossover}
\end{equation}
The second estimate is controlled by its lower endpoint,
\(z\sim z_t\). Thus, the \(1/t\) behavior is generated by the
neighborhood of the crossover and receives contributions from both
\(z\lesssim z_t\) and \(z\gtrsim z_t\). 

The same \(r=0\) coefficient also contains a logarithm, which comes
from a subleading term in the local expansion. Define
\begin{equation}
G_t=a(x,y)z+b(x,y)t,
\qquad
H_t=cz^2+d(x,y)tz,
\qquad
F_D=G_t+H_t.
\end{equation}
Where \(|H_t|\ll|G_t|\), the scalar integrand becomes
\begin{equation}
\frac{1}{F_D^2}
=
\frac{1}{G_t^2}
-\frac{2\bigl[cz^2+d(x,y)tz\bigr]}{G_t^3}
+\cdots.
\label{eq:nonzero-t-local-expansion}
\end{equation}
For \(z\lesssim z_t\), \(G_t\simeq b(x,y)t\). The integrals of both
correction terms over this range remain finite as \(t\to0\); this
range does not generate the logarithm. For \(z\gtrsim z_t\),
\(G_t\simeq a(x,y)z\), and the \(cz^2\) correction behaves as
\(1/z\). It therefore gives
\begin{align}
-2c\int_{z_t}^{z_0}\mathrm{d}z\,
\frac{z^2}{\bigl[a(x,y)z+b(x,y)t\bigr]^3}
&\sim
-\frac{2c}{a(x,y)^3}
\ln\!\left(\frac{z_0}{z_t}\right).
\label{eq:t-log}
\end{align}
Since \(z_t\propto |t|\) at fixed \(x,y\), this contains a logarithm of
\(|t|\). The causal branch is understood. The term proportional to
\(d(x,y)tz\) behaves as \(t/z^2\) above the crossover, and its integral
is of order \(t/z_t\sim t^0\); it does not produce an unsuppressed
\(\ln|t|\) term. The pole and logarithm therefore probe the same
triangle boundary but arise from different terms in the local
expansion.

For \(r=1\), the local integrand instead contains \(F_D^{-1}\). The
leading boundary form already yields a logarithm:
\begin{equation}
\int_0^{z_0}
\frac{\mathrm{d}z}{a(x,y)z+b(x,y)t}
=
\frac{1}{a(x,y)}
\ln\!\left[
\frac{a(x,y)z_0+b(x,y)t}{b(x,y)t}
\right].
\label{eq:r-one-leading-log}
\end{equation}
Its nonanalytic dependence as \(t\to0\) is logarithmic, with the
causal prescription understood.

Recall that at \(t=0\), the box endpoint is power divergent in
four dimensions. Analytic continuation in DR conceals this
divergence, which gives no \(1/\epsilon\) pole in the expansion
about \(d=4\). At the critical dimension \(d=6\), the same
endpoint is logarithmically divergent, and the analytically
continued endpoint integral has a pole. Keeping \(t\neq0\)
regulates the same boundary region and makes the \(1/t\) behavior
of the four-point coefficient functions explicit as \(t\to0\). After the
relevant diagrams are combined and the appropriate tensor structures
are projected, the residues are proportional to matrix elements of
the gluonic operators
\(F^{\mu\nu}F_{\mu\nu}\) and
\(F^{\mu\nu}\widetilde{F}_{\mu\nu}\). Thus, the boundary analysis
identifies the geometry that generates the anomaly-related pole
residues found in earlier work.

The Compton amplitude contains 22 distinct four-point coefficient
functions from the family in Eq.~\eqref{eq:D-index-set}. We focus on three representative coefficients to illustrate how their
nonanalytic dependence on \(t\) is governed by the powers of \(z\) in
the numerator and of \(F_D\) in the denominator of the parameter
representation. Define
\begin{equation}
\Delta_t
\equiv
(Q_a^2+s)(Q_b^2+s)+st,
\qquad
\lambda_t
\equiv
\lambda(-Q_a^2,-Q_b^2,t),
\end{equation}
where the K\"allen function
\begin{equation}
\lambda(x,y,z)=x^2+y^2+z^2-2xy-2xz-2yz.
\end{equation}
We begin with the scalar coefficient, \(r=0\) and \(n_3=0\):
\begin{equation}
D_0\left(
0,0,-Q_b^2,-Q_a^2;\,
t,s;\,
0,0,0,0
\right),
\end{equation}
which has the expansion
\begin{align}
D_0
={}&
\frac{1}{\epsilon^2 s t}
-\frac{\pi^2}{12s t}
-\frac{2}{s t}\,
 \operatorname{Li}_2\!\left(
 \frac{Q_a^2+s}{s}
 \right)
-\frac{2}{s t}\,
 \operatorname{Li}_2\!\left(
 \frac{Q_b^2+s}{s}
 \right)
\nonumber\\
&-\frac{1}{s t}
 \ln^2\!\left(\frac{t}{s}\right)
-\frac{1}{\epsilon s t}
 \ln\!\left(\frac{\mu^2}{Q_a^2}\right)
-\frac{1}{2s t}
 \ln^2\!\left(\frac{\mu^2}{Q_a^2}\right)
-\frac{1}{2s t}
 \ln^2\!\left(\frac{\mu^2}{Q_b^2}\right)
\nonumber\\
&+\frac{1}{s t}
 \ln^2\!\left(-\frac{\mu^2}{s}\right)
-\frac{1}{s t}
 \ln\!\left(\frac{\mu^2}{Q_b^2}\right)
 \ln\!\left(-\frac{\mu^2}{t}\right)
-\frac{1}{s t}
 \ln^2\!\left(-\frac{\mu^2}{t}\right)
+\frac{3}{2s t}
 \ln^2\!\left(-\frac{\mu^2}{t}\right)
\nonumber\\
&+\frac{1}{s t}
 \ln\!\left(\frac{\mu^2}{Q_a^2}\right)
 \left[
 \ln\!\left(\frac{\mu^2}{Q_b^2}\right)
 -\ln\!\left(-\frac{\mu^2}{t}\right)
 \right]
+\frac{1}{\epsilon s t}
 \left[
 -\ln\!\left(\frac{\mu^2}{Q_b^2}\right)
 +2\ln\!\left(-\frac{\mu^2}{s}\right)
 +\ln\!\left(-\frac{\mu^2}{t}\right)
 \right]
+\mathcal O(\epsilon).
\label{eq:D0-nonzero-t}
\end{align}
The origin of $1/t$ is
the \(z=0\) boundary identified above, where the scalar parameter
integrand has no numerator factor that vanishes with \(z\). The integrand therefore retains its strongest singular behavior at
the \(z=0\) boundary.
For comparison, consider the \(r=0\) coefficient with \(n_3=1\),
denoted \(D_3\). Its expansion is
\begin{align}
D_3
={}&
\frac{
 \operatorname{Li}_2\!\left(\dfrac{Q_a^2+s}{s}\right)
+\operatorname{Li}_2\!\left(\dfrac{Q_b^2+s}{s}\right)}
{\Delta_t}
+\frac{1}{2\Delta_t}\ln^2\!\left(\frac{t}{s}\right)
+\frac{1}{2\Delta_t}
 \ln\!\left(\frac{\mu^2}{Q_b^2}\right)
 \ln\!\left(-\frac{\mu^2}{t}\right)
-\frac{1}{2\Delta_t}
 \ln^2\!\left(-\frac{\mu^2}{t}\right)
\nonumber\\
&+\ln\!\left(\frac{\mu^2}{Q_a^2}\right)
 \left[
 -\frac{1}{2\Delta_t}\ln\!\left(\frac{\mu^2}{Q_b^2}\right)
 +\frac{1}{2\Delta_t}\ln\!\left(-\frac{\mu^2}{t}\right)
 \right]
-\frac{1}{2\Delta_t}
+(Q_a^2+Q_b^2+2s+t)\,\mathcal B(Q_a^2,Q_b^2,t),
\label{eq:D3-nonzero-t}
\end{align}
where the dilogarithm combination is
\begin{align}
\mathcal B(Q_a^2,Q_b^2,t)
\equiv \frac{1}{\sqrt{\lambda_t}}\Bigg\{&
-\operatorname{Li}_2\!\left(
 \frac{-Q_a^2-Q_b^2-t-\sqrt{\lambda_t}}
      {-Q_a^2-Q_b^2-t+\sqrt{\lambda_t}}
 \right)
+\operatorname{Li}_2\!\left(
 \frac{-Q_a^2-Q_b^2-t+\sqrt{\lambda_t}}
      {-Q_a^2-Q_b^2-t-\sqrt{\lambda_t}}
 \right)
\nonumber\\
&-\operatorname{Li}_2\!\left(
 \frac{Q_a^2-Q_b^2+t-\sqrt{\lambda_t}}
      {Q_a^2-Q_b^2+t+\sqrt{\lambda_t}}
 \right)
+\operatorname{Li}_2\!\left(
 \frac{Q_a^2-Q_b^2+t+\sqrt{\lambda_t}}
      {-Q_a^2+Q_b^2-t+\sqrt{\lambda_t}}
 \right)
\nonumber\\
&-\operatorname{Li}_2\!\left(
 \frac{-Q_a^2+Q_b^2+t-\sqrt{\lambda_t}}
      {-Q_a^2+Q_b^2+t+\sqrt{\lambda_t}}
 \right)
+\operatorname{Li}_2\!\left(
 \frac{-Q_a^2+Q_b^2+t+\sqrt{\lambda_t}}
      {-Q_a^2+Q_b^2+t-\sqrt{\lambda_t}}
 \right)
\Bigg\}.
\label{eq:B-dilog-combination}
\end{align}
This coefficient contains logarithmic dependence on \(t\), but not the
same \(1/t\) enhancement as \(D_0\). This is consistent with
Eq.~\eqref{eq:D-boundary-scaling}: for \(n_3=1\), the numerator supplies
a factor of \(z\), which vanishes on the \(z=0\) boundary and softens
the endpoint behavior.

Finally, the \(r=1,\ n_3=0\) example, denoted \(D_{00}\), has the
expansion
\begin{align}
D_{00}
={}&
\frac{s}{2\Delta_t}
 \operatorname{Li}_2\!\left(\frac{Q_a^2+s}{s}\right)
+\frac{s}{2\Delta_t}
 \operatorname{Li}_2\!\left(\frac{Q_b^2+s}{s}\right)
+\frac{s}{4\Delta_t}\ln^2\!\left(\frac{t}{s}\right)
+\frac{s}{4\Delta_t}
 \ln\!\left(\frac{\mu^2}{Q_a^2}\right)
 \ln\!\left(-\frac{\mu^2}{t}\right)
-\frac{s}{4\Delta_t}
 \ln^2\!\left(-\frac{\mu^2}{t}\right)
\nonumber\\
&+\ln\!\left(\frac{\mu^2}{Q_b^2}\right)
 \left[
 -\frac{s}{4\Delta_t}\ln\!\left(\frac{\mu^2}{Q_a^2}\right)
 +\frac{s}{4\Delta_t}\ln\!\left(-\frac{\mu^2}{t}\right)
 \right]
+\frac{
 2Q_a^2Q_b^2+Q_a^2s+Q_b^2s+st
}{4\Delta_t}
+\mathcal B(Q_a^2,Q_b^2,t).
\label{eq:D00-nonzero-t}
\end{align}
Here the reduced power of the denominator in the parameter
representation gives logarithmic dependence on \(t\) already at
leading order near the boundary.

To summarize, for \(r=1\), the parameter representation contains one fewer power of
\(F_D\) in the denominator. Near the \(z=0\) boundary, its leading
local behavior is therefore proportional to
\((a(x,y)z+b(x,y)t)^{-1}\). Integration over \(z\) produces
logarithmic dependence on \(t\). By contrast, when \(r=0\) and
\(n_3=1\), the numerator factor \(z\) suppresses the \(1/t\)
enhancement found for \(n_3=0\), although logarithmic dependence on
\(t\) may remain.
These three examples illustrate the distinct boundary behaviors
within the four-point coefficient family. The remaining coefficients can be classified by the
same parameter-space power counting.
\footnote{The poles displayed here pertain to individual
Passarino--Veltman coefficients. In the complete diagram, these
coefficients are multiplied by the tensor-reduction coefficients.
The \(1/\epsilon^2\) pole and the mixed \(1/\epsilon\)-logarithmic
terms cancel in the resulting sum.}

As discussed above, the propagator denominators determine the
location of the Landau surface, whereas the residue depends on
the numerator and on how the gluon polarizations are treated.
The DVCS analyses of
Refs.~\cite{Bhattacharya:2022xxw,Bhattacharya:2023wvy}
kept the gluon polarization vectors explicit and identified the
tensor structures
\begin{equation}
\begin{aligned}
(\epsilon_1\cdot\ell)(\epsilon_2^*\cdot\ell)
-\frac{1}{2}(\epsilon_1\cdot\epsilon_2^*)\ell^2
&\propto
\langle p_2|F^{\mu\nu}F_{\mu\nu}|p_1\rangle,
\\
2i\epsilon^{\epsilon_1\epsilon_2^*\ell p}
&\propto
\langle p_2|F^{\mu\nu}\widetilde F_{\mu\nu}|p_1\rangle .
\end{aligned}
\label{eq:dvcs_tensor_structures}
\end{equation}
Here \(\epsilon^{\epsilon_1\epsilon_2^*\ell p}\) denotes the
Levi-Civita contraction of the four indicated vectors, with
\(\epsilon_1=\epsilon(p_1)\) and \(\epsilon_2^*=\epsilon^*(p_2)\).
The first structure is associated with the trace anomaly and
the second with the chiral anomaly. In the loop amplitude,
the singular Passarino--Veltman coefficients identified above
generate \(1/t\) terms with these scalar and pseudoscalar
structures.

To see why keeping the gluon polarizations explicit matters,
consider the numerator tensor \(k^\alpha k^\beta\) (where $k$ is the loop momentum). Its
contraction with the polarization vectors gives
\begin{equation}
\epsilon_\alpha(p_1)\epsilon_\beta^*(p_2)k^\alpha k^\beta
=
\bigl(\epsilon(p_1)\cdot k\bigr)
\bigl(\epsilon^*(p_2)\cdot k\bigr).
\end{equation}
Tensor reduction can generate a term proportional to
\(\ell^\alpha\ell^\beta\). Thus, its contraction gives
\((\epsilon_1\cdot\ell)(\epsilon_2^*\cdot\ell)\), which enters
the scalar structure in Eq.~\eqref{eq:dvcs_tensor_structures}.

We now examine how the transverse projectors act on this
numerator tensor.  The symmetric projector \(g_\perp^{\alpha\beta}\) gives a scalar
that is invariant under rotations in the transverse plane:
\begin{equation}
g_{\perp\,\alpha\beta}k^\alpha k^\beta=k^2.
\end{equation}
Using Eq.~\eqref{eq:transverse-projector-numerator-identities},
we can express the resulting numerator in terms of propagator
denominators. The cancellations reorganize the box contribution
into bubble and triangle integrals and remaining box terms, none
of which generates the anomaly-related \(1/t\) pole.
Likewise, the antisymmetric projector annihilates this particular numerator
tensor by symmetry:
\begin{equation}
\epsilon_{\perp\,\alpha\beta}k^\alpha k^\beta=0.
\end{equation}
The other numerator terms can contribute to the projected
antisymmetric amplitude and give no \(1/t\) pole in the
calculation considered here.

Thus, keeping the gluon polarization vectors explicit allows us to
track the tensor structures that carry the anomaly-related
residues (see Eq.~\eqref{eq:dvcs_tensor_structures}). A transverse projector can suppress these residues by changing
the numerator and reducing the box contribution to a linear
combination of bubble, triangle, and box integrals that yield
no anomaly-related pole. We emphasize that the suppression concerns the residue
in the projected amplitude. The original box denominators still
satisfy the Landau conditions on the same boundary.

\subsection{Nonzero quark mass (\(m\neq0) \)}
We next regulate the same boundary by assigning a common mass \(m\) to
the four internal propagators and setting \(t=0\). The resulting
mass-regulated expressions also describe the leading behavior in the
hierarchy \(\lvert t\rvert\ll m^{2}\ll Q^{2}\). The case with both
\(t\) and \(m\) nonzero will be discussed briefly later in this subsection; a systematic analysis of
the joint limit is deferred to a separate work. 

The tensor-reduced
amplitude then contains
\begin{equation}
D_I\left(
0,0,-Q_b^2,-Q_a^2;
0,s;
m^2,m^2,m^2,m^2
\right),
\qquad I\in\mathcal I_D.
\label{eq:mass-regulated-D-family}
\end{equation}
Because the Feynman parameters satisfy \(\sum_i\alpha_i=1\), the mass
contribution to the denominator polynomial is
\(m^2\sum_i\alpha_i=m^2\). Hence
\begin{equation}
F_D^{(m)}(x,y,z)
=
m^2+a(x,y)z+cz^2,
\label{eq:FD-mass-regulator}
\end{equation}
where \(a(x,y)\) is defined in Eq.~\eqref{eq:a-xy}, and \(c\) is given
in Eq.~\eqref{eq:small-t-bcd}. In the massless limit this polynomial
vanishes on the \(z=0\) face associated with the reduced triangle in
Fig.~\ref{fig:z-zero-reduction}. A nonzero quark mass lifts the boundary value of the polynomial
from zero to \(m^2\), giving \(F_D^{(m)}(x,y,0)=m^2\).
At generic fixed \(x,y\), the mass term competes with the term linear
in \(z\) at
\begin{equation}
z\sim z_m,
\qquad
z_m\equiv\frac{m^2}{|a(x,y)|}.
\label{eq:zm-crossover}
\end{equation}
As in the finite-\(t\) case, choose a fixed, small upper limit \(z_0\)
for the boundary region. This does not change the lower endpoint of
the original integral, which remains \(z=0\). The split at \(z_m\)
only separates the ranges in which \(m^2\) and \(a(x,y)z\) dominate.

For \(r=0\), the scalar coefficient has a local integrand
proportional to \(\bigl(F_D^{(m)}\bigr)^{-2}\). Keeping the leading
terms gives
\begin{align}
I_m
&\equiv
\int_0^{z_0}
\frac{\mathrm{d}z}{\bigl[a(x,y)z+m^2\bigr]^2}
\nonumber\\
&=
\frac{1}{a(x,y)m^2}
-\frac{1}{
a(x,y)\bigl[a(x,y)z_0+m^2\bigr]}.
\label{eq:mass-leading-pole}
\end{align}
For generic \(a(x,y)\neq0\), the second term remains finite as
\(m^2\to0\), while the first is proportional to \(1/m^2\).
Both sides of \(z_m\) contribute at this inverse-mass order. Below the
crossover,
\begin{equation}
\int_0^{z_m}
\frac{\mathrm{d}z}{\bigl[a(x,y)z+m^2\bigr]^2}
\sim
\int_0^{z_m}\frac{\mathrm{d}z}{m^4}
=
\frac{z_m}{m^4}
\sim
\frac{1}{|a(x,y)|m^2}.
\label{eq:mass-below-crossover}
\end{equation}
Above the crossover, the denominator is approximately \(a(x,y)z\):
\begin{equation}
\int_{z_m}^{z_0}
\frac{\mathrm{d}z}{\bigl[a(x,y)z+m^2\bigr]^2}
\sim
\frac{1}{a(x,y)^2}
\left(\frac{1}{z_m}-\frac{1}{z_0}\right)
\sim
\frac{1}{|a(x,y)|m^2}.
\label{eq:mass-above-crossover}
\end{equation}
This second integral is dominated by its lower endpoint,
\(z\sim z_m\). Thus, the inverse-mass behavior receives contributions
from both \(z\lesssim z_m\) and \(z\gtrsim z_m\). As we will show in the next section (Sec.~\ref{sec:complete-results}), after summing the contributing diagrams and projecting onto the
relevant tensor structures, the \(1/m^2\) terms do not cancel. We therefore find an inverse-mass-squared contribution in the
complete amplitude. To our knowledge, this behavior
has not previously been reported for these amplitudes. The
parameter analysis traces this term to the neighborhood of the
\(z=0\) boundary associated with the reduced triangle.
The origin of this contribution at the level of the full QCD operator
expansion is not addressed here. In particular, we do not identify it
with a specific operator matrix element or assign it an independent
operator interpretation. We regard the result as a feature of the
calculated amplitude whose interpretation requires further study.

The \(r=0\) coefficient also contains a logarithm from the first
correction to the local denominator. Write
\begin{equation}
F_D^{(m)}=G_m+H_m,
\qquad
G_m=a(x,y)z+m^2,
\qquad
H_m=cz^2.
\end{equation}
The expansion gives
\begin{equation}
\frac{1}{\bigl(F_D^{(m)}\bigr)^2}
=
\frac{1}{G_m^2}
-\frac{2cz^2}{G_m^3}
+\cdots.
\label{eq:mass-local-expansion}
\end{equation}
For \(z\lesssim z_m\), the correction is proportional to
\(z^2/m^6\), and its integral remains finite as \(m^2\to0\). For
\(z\gtrsim z_m\), it behaves as \(1/z\), giving
\begin{align}
-2c\int_{z_m}^{z_0}\mathrm{d}z\,
\frac{z^2}{\bigl[a(x,y)z+m^2\bigr]^3}
&\sim
-\frac{2c}{a(x,y)^3}
\ln\!\left(\frac{z_0}{z_m}\right).
\label{eq:mass-log}
\end{align}
Since \(z_m\propto m^2\), this contains a logarithm of \(m^2\). The
logarithm accumulates over the interval above the crossover; it is not
generated by the region \(z\ll z_m\).

For \(r=1\), the local integrand instead contains
\(\bigl(F_D^{(m)}\bigr)^{-1}\). The leading boundary form already gives
\begin{equation}
\int_0^{z_0}
\frac{\mathrm{d}z}{a(x,y)z+m^2}
=
\frac{1}{a(x,y)}
\ln\!\left[
\frac{a(x,y)z_0+m^2}{m^2}
\right],
\label{eq:mass-r-one-log}
\end{equation}
and hence a logarithm as \(m^2\to0\).

The comparison with nonzero momentum transfer is direct. At generic
fixed \(x,y\), the two regulators enter the leading local denominator
as
\begin{equation}
a(x,y)z+b(x,y)t
\qquad\longleftrightarrow\qquad
a(x,y)z+m^2.
\end{equation}
They set different crossover scales, \(z_t=|b(x,y)t/a(x,y)|\) and
\(z_m=m^2/|a(x,y)|\), but regulate the same endpoint region near \(z=0\).
For \(r=0\), both sides of the crossover contribute to the
inverse-power behavior, \(1/t\) or \(1/m^2\), while the logarithm
comes from the correction that produces a \(1/z\) integrand above the
crossover. For \(r=1\), the leading local integrand already behaves
as \(1/z\) above the crossover and therefore produces a logarithm.

In the same spirit as in the previous subsection, we now examine the
same four-point coefficient functions with a common nonzero internal
mass \(m\), setting \(t=0\). For compactness, define
\begin{align}
R_a&\equiv \sqrt{Q_a^2(4m^2+Q_a^2)},&
R_b&\equiv \sqrt{Q_b^2(4m^2+Q_b^2)},&
R_s&\equiv \sqrt{s(s-4m^2)},\\
L_a&\equiv
\ln\!\left(
\frac{2m^2+Q_a^2+R_a}{2m^2}
\right),&
L_b&\equiv
\ln\!\left(
\frac{2m^2+Q_b^2+R_b}{2m^2}
\right),&
L_s&\equiv
\ln\!\left(
\frac{2m^2-s+R_s}{2m^2}
\right).
\label{eq:mass-regulated-logs}
\end{align}
We first consider the scalar coefficient, \(r=0\) and \(n_3=0\):
\begin{equation}
D_0^{(m)}
\equiv
D_0\left(
0,0,-Q_b^2,-Q_a^2;\,
0,s;\,
m^2,m^2,m^2,m^2
\right).
\end{equation}
Its mass dependence is
\begin{align}
D_0^{(m)}
={}&
\frac{R_a L_a}
{m^2(Q_a^2-Q_b^2)(Q_a^2+s)}
+\frac{R_b L_b}
{m^2(-Q_a^2+Q_b^2)(Q_b^2+s)}
+\frac{R_s L_s}
{m^2(Q_a^2+s)(Q_b^2+s)}.
\label{eq:D0-mass-regulated}
\end{align}
The explicit factor \(1/m^2\) displays the power enhancement in this
coefficient\footnote{Although the individual coefficient functions contain logarithms of
\(m^2\), these logarithms cancel in their sum. The complete result
therefore retains a \(1/m^2\) contribution without a logarithmic
enhancement.}. 
For comparison, the \(r=0,\ n_3=1\) coefficient is
\begin{align}
D_3^{(m)}
={}&
-\frac{L_a^2+L_b^2}
{2(Q_a^2-Q_b^2)(Q_a^2+s)}
+\frac{-L_b^2+L_s^2}
{2(Q_b^2+s)(Q_a^2+s)}.
\label{eq:D3-mass-regulated}
\end{align}
This expression contains logarithms of the mass-regulated invariants,
but no explicit \(1/m^2\) prefactor. The factor of \(z\) in its
parameter numerator vanishes at the \(z=0\) boundary and softens the
power enhancement found for \(D_0^{(m)}\).

The \(r=1,\ n_3=0\) coefficient provides a third example:
\begin{align}
D_{00}^{(m)}
={}&
\frac{R_a L_a}
{(Q_a^2-Q_b^2)(Q_a^2+s)}
+\frac{R_b L_b}
{(-Q_a^2+Q_b^2)(Q_b^2+s)} +\frac{R_sL_s}{(Q_a^2+s)(Q_b^2+s)}
\nonumber\\
&+\frac{Q_a^2}{2(Q_a^2+s)}
\left[
-\frac{L_a^2}{2(Q_a^2-Q_b^2)}
+\frac{L_b^2}{2(Q_a^2-Q_b^2)}
\right]
+\frac{s}{2(Q_a^2+s)}
\left[
-\frac{L_b^2}{2(Q_b^2+s)}
+\frac{L_s^2}{2(Q_b^2+s)}
\right].
\label{eq:D00-mass-regulated}
\end{align}
Like \(D_3^{(m)}\), this coefficient has logarithmic mass dependence
but no explicit \(1/m^2\) prefactor.

To summarize, introducing a nonzero mass exposes the same
\(z=0\) triangle boundary in the four-point coefficient functions that
is regulated by nonzero \(t\). After the relevant contributions are
combined, this boundary produces a \(1/m^2\) term. To repeat again, in the massless,
zero-transfer limit, the corresponding triangle vanishes in DR at \(d=4\) by analytic
continuation. This vanishing does not remove the underlying
power-divergent region. A cutoff regulator can make its power behavior
explicit, while in DR the same region becomes
logarithmically divergent at the critical dimension \(d=6\).

Note that in the preceding subsections, we considered the cases in which \(t\)
or \(m\) separately regulates the boundary. We can also
obtain the exact scalar coefficient with both regulators nonzero.
The exact expression involves square roots and continued
dilogarithms. We present it below in its unexpanded form for
illustration:
\begin{equation}
D_0\left(
0,0,-Q_b^2,-Q_a^2;\,
t,s;\,
m^2,m^2,m^2,m^2
\right).
\end{equation}
We use the abbreviations
\begin{equation}
M\equiv m^2,\qquad
A\equiv Q_a^2,\qquad
B\equiv Q_b^2,\qquad
K\equiv 2M+A+\sqrt{A(A+4M)},
\end{equation}
and
\begin{align}
U&\equiv 2M(2M+B)-(2M-s)(2M-t),\\
V&\equiv \frac{K}{2}(B+s)+Mt,\\
W&\equiv-\frac{2M(B+s)}{K}-t,\\
\Theta_X&\equiv U+2VX,\qquad
\chi\equiv\frac{K}{2}-\frac{2M^2}{K}.
\end{align}
The roots in the outer sum are specified by
\begin{equation}
P(X)\equiv VX^2+UX-MW=0.
\end{equation}
Here \(\mathscr D(u,v)\) and \(\mathscr L(u,v)\) denote the
branch-prescribed two-argument dilogarithm and logarithm, respectively,
in the convention of the loop-integral result. We write
\(\mathscr D_{\rm c}(\{\cdots\},\{\cdots\})\) for its continued
dilogarithm. Define
\begin{align}
R_X&\equiv
\frac{M(1+X^2)+X(2M+B)}
     {M(1+X^2)+X(2M-s)},\\
S_X&\equiv
\frac{
2M+X\chi-\dfrac{2M(2M+B)}{K}-t
}{
2M+X\chi-\dfrac{2M(2M-s)}{K}
},\\
J_X&\equiv
-\mathscr L\!\left(-X,\frac{W}{\Theta_X}\right)
+\mathscr L\!\left(-\frac{XK}{2M},
                   \frac{KW}{2M\Theta_X}\right)
-\ln\!\left(\frac{K}{2M}\right).
\end{align}
The inner sums are
\begin{align}
\mathcal R_B(X)
&\equiv
\sum_{\substack{r:\,Mr^2+(2M+B)r+M=0}}
\mathscr D_{\rm c}\!\left(
\left\{-\frac1r,-\frac Mr+Mr\right\},
\left\{-X,\frac{W}{2M\Theta_X}\right\}
\right),\\
\mathcal R_s(X)
&\equiv
\sum_{\substack{r:\,Mr^2+(2M-s)r+M=0}}
\mathscr D_{\rm c}\!\left(
\left\{-\frac1r,-\frac Mr+Mr\right\},
\left\{-X,\frac{W}{2M\Theta_X}\right\}
\right),\\
\mathcal R_t(X)
&\equiv
\sum_{\substack{r:\,Mr^2+(2M-t)r+M=0}}
\mathscr D_{\rm c}\!\left(
\left\{-\frac1r,-\frac Mr+Mr\right\},
\left\{-\frac{XK}{2M},\frac{KW}{2M\Theta_X}\right\}
\right).
\end{align}
Then
\begin{align}
D_0
&=
\sum_{\substack{X:\,P(X)=0}}
\frac{\Phi(X)}{\Theta_X},
\label{eq:D0-both-regulators}
\\
\Phi(X)
&\equiv
\mathscr D\!\left(
1+\frac{XK}{2M},
-\frac{XK}{2M}+\frac{KW}{2M\Theta_X}
\right)
\quad+
\mathscr D\!\left(
1+\frac{XK}{2M},
\frac{XK}{2M}+\frac{KW}{2M\Theta_X}
\right)
\nonumber\\
&\quad -\mathscr L(R_X,S_X)\,J_X +\mathcal R_B(X)-\mathcal R_s(X)-\mathcal R_t(X).
\label{eq:D0-both-regulators-summand}
\end{align}
The branch prescriptions in \(\mathscr D\), \(\mathscr L\), and
\(\mathscr D_{\rm c}\) are inherited from the loop integral. Although the expression is exact for nonzero \(t\) and \(m\), extracting
its asymptotic behavior requires careful control of the branches and
the order of limits. The two relevant hierarchies suggest the
following leading behavior for the complete amplitude:
\begin{equation}
\mathcal A(t,m^2)\sim
\begin{cases}
\dfrac{\mathcal R}{t}
+\mathcal O\!\left(\dfrac{m^2}{|t|}\right),
& m^2\ll |t|\ll Q^2,\\[3mm]
\dfrac{\mathcal R}{m^2}
+\mathcal O\!\left(\dfrac{|t|}{m^2}\right),
& |t|\ll m^2\ll Q^2.
\end{cases}
\label{eq:joint-regulator-hierarchies}
\end{equation}
Here \(\mathcal R\) denotes the common residue (chiral or trace anomaly) in the
complete amplitude. Thus, the leading singular contribution is
associated with the same anomaly-related tensor structure in either
hierarchy. A controlled derivation of these limits,
including the branch choices and their relation to dimensional
regularization, will be presented in a separate work. For clarity, we
focus here on the cases in which \(t\) or \(m\) separately regulates
the boundary.

\section{Complete one-loop results in DDVCS}
\label{sec:complete-results}
We now combine the box, cat's-eye, and hourglass contributions
shown in Fig.~\ref{fig:three-gluon-diagrams} and present the
complete one-loop results using three complementary infrared
prescriptions: DR, nonzero momentum
transfer (\(t\neq0\)), and a finite quark mass (\(m\neq0\)).
The comparison shows how the anomaly-related terms and the
infrared singularities associated with evolution are manifested
in each prescription.
The regulator dependence of the complete one-loop result can be
summarized schematically as follows:
\begin{itemize}
\item \textbf{DR at \(t=m=0\):}
\begin{equation}
\mathcal R_{\rm bdry}\,
\underbrace{\mathcal P_{\rm lin}^{\rm DR}}_{=\,0\ \text{at }d=4}
+\frac{\mathcal K_{\rm evol}}{\epsilon_{\rm IR}}
+\mathcal T_{\rm rem}^{\mathrm{DR}}.
\label{eq:complete-result-dr-schematic}
\end{equation}
Here \(\mathcal P_{\rm lin}^{\rm DR}\) is a bookkeeping symbol
for the power-divergent boundary sector, which diverges as
\(1/\delta\) at \(d=4\) with a lower endpoint cutoff \(\delta\).
We set this symbol to zero in the DR result because analytic
continuation yields no \(1/\epsilon\) pole from this sector
in the expansion about \(d=4\). This does not mean that the
underlying singular region is absent: the \(1/\delta\)
divergence is explicit in the cutoff calculation. Its absence
from the pole part at \(d=4\) is a consequence of analytic
continuation. As discussed in
Sec.~\ref{subsec:dimensional-regularization}, the same endpoint
diverges logarithmically at the critical dimension \(d=6\),
where the analytically continued integral has a pole at
\(\epsilon=-1\).

\item \textbf{Nonzero momentum transfer, \(t\neq0\):}
\begin{equation}
\frac{\mathcal R_t}{t}
+\mathcal K_{\rm evol}
 \ln\!\left(\frac{Q^2}{-t}\right)
+\mathcal T_{\rm rem}^{t}.
\label{eq:complete-result-t-schematic}
\end{equation}

\item \textbf{Nonzero quark mass, \(m\neq0\):}
\begin{equation}
\frac{\mathcal R_m}{m^2}
+\mathcal K_{\rm evol}
 \ln\!\left(\frac{Q^2}{m^2}\right)
+\mathcal T_{\rm rem}^{m}.
\label{eq:complete-result-m-schematic}
\end{equation}
\end{itemize}
A few comments are in order. First, in the nonzero-\(t\) calculation, we do not impose the lightlike relation
\(\ell^\mu=-2\xi p^\mu\)~\cite{Bhattacharya:2022xxw,Bhattacharya:2023wvy}, which would set \(t=0\).
In the mass-regulated calculation as well we do not use
\(\ell^\mu =-2\xi p^\mu\). This choice allows us to retain the polarization-dependent tensor
structures that carry the anomaly-related contribution (see Eq.~\eqref{eq:dvcs_tensor_structures}).
Second, we emphasize again that the symbol \(\mathcal P_{\rm lin}^{\rm DR}=0\) records how
DR treats the boundary contribution.
It does not imply that the underlying power-divergent region is absent.
Third, the \(1/t\) and \(1/m^2\) terms expose the anomaly-related
residues \(\mathcal R_t\) and \(\mathcal R_m\), respectively.
The gluonic tensor structures identified in
Eq.~\eqref{eq:dvcs_tensor_structures} allow us to write the
residues of the \(1/t\) and \(1/m^2\) terms schematically as
\begin{equation}
\mathcal R_\rho^{X}
=
C_\rho^{X}\,
\langle p_2|\mathcal O_X|p_1\rangle,
\qquad
\rho=t,m,\quad X=A,S,
\label{eq:residue-operator-structures}
\end{equation}
where \(A\) and \(S\) denote the antisymmetric (polarized) and
symmetric (unpolarized) amplitudes, respectively, and
\(\mathcal O_A=F^{\mu\nu}\widetilde F_{\mu\nu}\) and
\(\mathcal O_S=F^{\mu\nu}F_{\mu\nu}\).
Within each polarization channel (polarized or unpolarized),
the \(1/t\) and \(1/m^2\) residues share the same gluonic
operator structure, while the coefficient functions
\(C_t^X\) and \(C_m^X\) may differ. This shared operator structure is consistent with
the common origin of the two inverse-power terms.
We leave a comparison of the \(x\)-integrals of the coefficient
functions, along the lines of
Refs.~\cite{Bhattacharya:2022xxw,Bhattacharya:2023wvy},
to future work.
Finally, the infrared pole \(1/\epsilon_{\rm IR}\) and the
regulator-dependent logarithms share the same evolution-kernel
coefficient, \(\mathcal K_{\rm evol}\). Moreover, each \(\mathcal T_{\rm rem}\) denotes the remainder finite in the
corresponding regulator limit; its dependence on the polarization
channel and external kinematics is implicit.

Note that the ultraviolet-pole cancellations follow a different, channel-dependent
pattern. In the symmetric amplitude, the ultraviolet poles from the
box and hourglass diagrams cancel against the pole from the cat's-eye
diagram. In the antisymmetric amplitude, the ultraviolet poles from
the box and hourglass diagrams cancel each other. We are now ready to present the complete results explicitly.

\subsection{Polarized case}
For the polarized amplitude, the evolution kernel is
\begin{align}
\widetilde{\mathcal K}_{\rm evol}(\hat{x},\hat{\xi})
={}&
\frac{2\hat{x}-1-\hat{\xi}^{2}}
     {(1-\hat{\xi}^{2})^{2}}
\ln\!\left(\frac{\hat{x}-1}{\hat{x}}\right)
-
2\frac{\hat{x}-\hat{\xi}}
        {(1-\hat{\xi}^{2})^{2}}
\ln\!\left(\frac{\hat{x}-\hat{\xi}}{\hat{x}}\right)
-\bigl(\hat{x}\to-\hat{x}\bigr).
\label{eq:evolution-kernel-polarized}
\end{align}
This expression agrees with the previously reported polarized
result~\cite{Belitsky:1997rh,Ji:1997nk,Pire:2011st,Bhattacharya:2022xxw,Bhattacharya:2023wvy}. The same kernel multiplies the infrared pole \(1/\epsilon_{\rm IR}\) in dimensional
regularization and the regulator logarithms in the nonzero-\(t\) and
nonzero-\(m\) calculations.

In the notation of Ref.~\cite{Bhattacharya:2023wvy},
the coefficient of the anomaly-related \(1/t\) term is given
by:
\begin{equation}
\begin{aligned}
\mathcal R_t^A (\hat{x},\hat{\xi})
&\sim
\frac{8\langle p_2|F^{\mu\nu}\widetilde F_{\mu\nu}|p_1\rangle }{(1-\hat{\xi}^{\,2})}
\Biggl \{
(1-\hat{x})\ln\!\left(\frac{\hat{x}-1}{x}\right)
+(\hat{x}-\hat{\xi})
 \ln\!\left(\frac{\hat{x}-\hat{\xi}}{\hat{x}}\right)
-(\hat{x}\to-\hat{x})
\Biggr \}.
\end{aligned}
\label{eq:t-regulated-axial-residue}
\end{equation}
With the mass regulator, the coefficient of the \(1/m^2\) term is given by:
\begin{align}
\mathcal R_m^{A}(\hat{x},\hat{\xi})
\sim {}&
\frac{4\,\langle p_2|F^{\mu\nu}\widetilde F_{\mu\nu}|p_1\rangle}
{3\hat{\xi}(\hat{\xi}^{2}-1)^2}
\Bigg\{
-4\hat{x}\hat{\xi}
+4\hat{x}\hat{\xi}^{3}
+2\hat{\xi}(\hat{x}-1)^3\ln(\hat{x}-1)
+2\hat{\xi}(\hat{x}+1)^3\ln(\hat{x}+1)
\nonumber\\
&\quad
+\Bigl[
3\hat{x}^{2}-6\hat{x}\hat{\xi}-2\hat{x}^{3}\hat{\xi}
+3\hat{\xi}^{2}+3\hat{x}^{2}\hat{\xi}^{2}-\hat{\xi}^{4}
\Bigr]\ln(\hat{x}-\hat{\xi})
\nonumber\\
&\quad
+\Bigl[
-3\hat{x}^{2}-6\hat{x}\hat{\xi}-2\hat{x}^{3}\hat{\xi}
-3\hat{\xi}^{2}-3\hat{x}^{2}\hat{\xi}^{2}+\hat{\xi}^{4}
\Bigr]\ln(\hat{x}+\hat{\xi})
\Bigg\}.
\label{eq:antisymmetric-mass-residue}
\end{align}
Thus, the coefficients of the \(1/t\) and \(1/m^2\) terms multiply
the same pseudoscalar gluonic tensor structure, which appears in
the matrix element of \(F^{\mu\nu}\widetilde F_{\mu\nu}\) and is
associated with the chiral anomaly. 
As discussed above, keeping the gluon polarization vectors
explicit makes this structure manifest in the residue.
Contracting with a transverse projector can reorganize the
numerator and suppress the residue, while the original propagator
denominators still satisfy the Landau conditions on the same
boundary. Both regulators, \(t\neq0\)
and \(m\neq0\), therefore, expose the same Landau boundary, although the associated \((\hat{x},\hat{\xi})\)-dependent coefficient functions may differ.

We now turn to the finite terms. The DR result
contains the following coefficient function:
\begin{align}
\mathcal T_{\rm rem}^{A, {\rm DR}}(\hat{x},\hat{\xi})
\sim{}&
-\frac{4Q^{2}}
{\hat{x}(\hat{\xi}^{2}-1)}
\Bigg\{
-4(\hat{x}-\hat{\xi}^{2})
 \ln\!\left(\frac{\hat{x}}{\hat{x}-1}\right)
+(-1+2\hat{x}-\hat{\xi}^{2})
 \ln^{2}\!\left(\frac{\hat{x}}{\hat{x}-1}\right)
\nonumber\\
&\quad
-4(\hat{x}+\hat{\xi}^{2})
 \ln\!\left(\frac{\hat{x}}{1+\hat{x}}\right)
+(1+2\hat{x}+\hat{\xi}^{2})
 \ln^{2}\!\left(\frac{\hat{x}}{1+\hat{x}}\right)
\nonumber\\
&\quad
+4(\hat{x}-\hat{\xi})
 \ln\!\left(\frac{\hat{x}}{\hat{x}-\hat{\xi}}\right)
-2(\hat{x}-\hat{\xi})
 \ln^{2}\!\left(\frac{\hat{x}}{\hat{x}-\hat{\xi}}\right)
\nonumber\\
&\quad
-2(\hat{x}+\hat{\xi})
\left[
-2+\ln\!\left(\frac{\hat{x}}{\hat{x}+\hat{\xi}}\right)
\right]
\ln\!\left(\frac{\hat{x}}{\hat{x}+\hat{\xi}}\right)
\Bigg\}.
\label{eq:polarized-DR-remainder}
\end{align}
For \(t\neq0\), the finite term in the notation of
Ref.~\cite{Bhattacharya:2023wvy} reads:
\begin{equation}
\begin{aligned}
\mathcal T_{\rm rem}^{A,t}(\hat{x},\hat{\xi})
\sim \frac{1}{(1-\hat{\xi}^{\,2})^2}
\Biggl\{&
(2\hat{x}-1-\hat{\xi}^{\,2})
\left[
  -\ln\!\frac{\hat{x}-1}{\hat{x}}
  +\frac{1}{2}\ln^2\!\frac{\hat{x}-1}{\hat{x}}
\right]
\\
&+2(\hat{x}-\hat{\xi})\ln\!\frac{\hat{x}-\hat{\xi}}{\hat{x}}
 +\hat{\xi}\ln^2\!\frac{\hat{x}-\hat{\xi}}{\hat{x}} -\hat{x}\ln\!\frac{\hat{x}-\hat{\xi}}{\hat{x}}\,
          \ln\!\frac{\hat{x}+\hat{\xi}}{\hat{x}}
\\
&
 +2\hat{\xi}\operatorname{Li}_2\!\left(
   \frac{-2\hat{\xi}}{\hat{x}-\hat{\xi}}\right)
+(2\hat{x}-1-\hat{\xi}^{\,2})
\left[
  \operatorname{Li}_2\!\left(\frac{1-\hat{\xi}}{1-\hat{x}}\right)
 +\operatorname{Li}_2\!\left(\frac{1+\hat{\xi}}{1-\hat{x}}\right)
\right]-(\hat{x}\to-\hat{x})\Biggr\}.
\end{aligned}
\label{eq:t-regulated-polarized-finite}
\end{equation}
Moving onto the $m \neq 0$ case: 
\begin{align}
\mathcal T_{\rm rem}^{A,m}(\hat{x},\hat{\xi})
\sim{}&
\frac{4}{\hat{\xi}^{2}-1}
\Bigg\{
(1-2\hat{x}+\hat{\xi}^{2})\ln^{2}(\hat{x}-1)
+\ln(\hat{x}-1)
\Bigl[
-6+8\hat{x}-2\hat{\xi}^{2}
+(-2+4\hat{x}-2\hat{\xi}^{2})\ln\hat{x}
\Bigr]
\nonumber\\
&\quad
+2\Bigl[
3+4\hat{x}+\hat{\xi}^{2}
+(1+2\hat{x}+\hat{\xi}^{2})\ln\hat{x}
\Bigr]\ln(1+\hat{x})
-(1+2\hat{x}+\hat{\xi}^{2})\ln^{2}(1+\hat{x})
\nonumber\\
&\quad
-4(\hat{x}-\hat{\xi})(2+\ln\hat{x})
 \ln(\hat{x}-\hat{\xi})
+2(\hat{x}-\hat{\xi})\ln^{2}(\hat{x}-\hat{\xi})
\nonumber\\
&\quad
-2(\hat{x}+\hat{\xi})
\Bigl[
4+2\ln\hat{x}-\ln(\hat{x}+\hat{\xi})
\Bigr]\ln(\hat{x}+\hat{\xi})
\Bigg\}.
\label{eq:antisymmetric-mass-remainder}
\end{align}

\subsection{Unpolarized case}
For the unpolarized amplitude, the evolution kernel is
\begin{align}
\mathcal K_{\rm evol}(\hat{x},\hat{\xi})
={}&
\frac{1-2\hat{x}+2\hat{x}^{2}-\hat{\xi}^{2}}
     {(1-\hat{\xi}^{2})^{2}}
\ln\!\left(\frac{\hat{x}-1}{\hat{x}}\right)
+
\frac{(\hat{x}-\hat{\xi})
      (1-2\hat{\xi}\hat{x}-\hat{\xi}^{2})}
     {\hat{\xi}(1-\hat{\xi}^{2})^{2}}
\ln\!\left(\frac{\hat{x}-\hat{\xi}}{\hat{x}}\right)
+\bigl(\hat{x}\to-\hat{x}\bigr).
\label{eq:evolution-kernel-unpolarized}
\end{align}
This expression agrees with the previously reported unpolarized
result~\cite{Belitsky:1997rh,Ji:1997nk,Pire:2011st,Bhattacharya:2022xxw,Bhattacharya:2023wvy}. As in the polarized case, the same kernel multiplies the
infrared pole \(1/\epsilon_{\rm IR}\)  and the regulator logarithms in all three calculations.
In each channel, this agreement across the three regularizations
provides a cross-check of the complete one-loop calculation.

In the notation of Ref.~\cite{Bhattacharya:2023wvy}, the coeffcient of the anomaly-related $1/t$ is given by:
\begin{equation}
\begin{aligned}
\mathcal R_t^S (\hat{x},\hat{\xi})
&\sim
2\langle p_2|F^{\mu\nu}F_{\mu\nu}|p_1\rangle
\Biggl\{
1+\frac{1}{1-\hat{\xi}^{\,2}}
\Biggl[
\hat{x}(1-\hat{x})
\ln\!\left(\frac{\hat{x}-1}{\hat{x}}\right)
+\hat{x}(\hat{x}-\hat{\xi})
\ln\!\left(\frac{\hat{x}-\hat{\xi}}{\hat{x}}\right)
+(\hat{x}\to-\hat{x})
\Biggr]
\Biggr\}.
\end{aligned}
\label{eq:t-regulated-trace-residue}
\end{equation}
On the other hand, with the mass regulator, the coefficient of the $1/m^2$ term reads:
\begin{align}
\mathcal R_m^{S}(\hat{x},\hat{\xi})
\sim {}&
-\frac{8\,\langle p_2|F^{\mu\nu}F_{\mu\nu}|p_1\rangle}
{3\hat{\xi}(\hat{\xi}^{2}-1)^2}
\Bigg\{
\hat{\xi}+4\hat{x}^{2}\hat{\xi}
-2\hat{\xi}^{3}-4\hat{x}^{2}\hat{\xi}^{3}+\hat{\xi}^{5}
-2\hat{x}\hat{\xi}(\hat{x}-1)^3\ln(\hat{x}-1)
-2\hat{x}\hat{\xi}(\hat{x}+1)^3\ln(\hat{x}+1)
\nonumber\\
&\quad
+\Bigl[
-3\hat{x}^{3}+6\hat{x}^{2}\hat{\xi}+2\hat{x}^{4}\hat{\xi}
-3\hat{x}\hat{\xi}^{2}-3\hat{x}^{3}\hat{\xi}^{2}
+\hat{x}\hat{\xi}^{4}
\Bigr]\ln(\hat{x}-\hat{\xi})
\nonumber\\
&\quad
+\Bigl[
3\hat{x}^{3}+6\hat{x}^{2}\hat{\xi}+2\hat{x}^{4}\hat{\xi}
+3\hat{x}\hat{\xi}^{2}+3\hat{x}^{3}\hat{\xi}^{2}
-\hat{x}\hat{\xi}^{4}
\Bigr]\ln(\hat{x}+\hat{\xi})
\Bigg\}.
\label{eq:symmetric-mass-residue}
\end{align}
The coefficients of the \(1/t\) and \(1/m^2\) terms multiply
the same scalar gluonic tensor structure, which appears in the
matrix element of \(F^{\mu\nu}F_{\mu\nu}\) and is associated with
the trace anomaly. Keeping the gluon polarization vectors explicit
makes this structure manifest in the residue. Setting
\(\ell^\mu=-2\xi p^\mu\) with \(p^2=0\) imposes \(t=\ell^2=0\)
from the outset, obscuring the \(1/t\) or $1/m^2$ behavior. 
As in the polarized case, contraction with a transverse projector
can suppress the residue through its effect on the numerator.
The Landau conditions continue to select the same boundary of the
box integral. To summarize, both regulators expose this boundary, although the
coefficient functions may differ.

Finally, we turn to the finite terms. The DR result contains the following coefficient function:
\begin{align}
\mathcal T_{\rm rem}^{S, {\rm DR}}(\hat{x},\hat{\xi})
\sim{}&
\frac{8}{\hat{\xi}(\hat{\xi}^{2}-1)}
\Bigg\{
-2(\hat{\xi}+\hat{\xi}^{3})
 \ln\!\left(\frac{\hat{x}}{\hat{x}-1}\right)
+\hat{\xi}(-1+2\hat{x}-2\hat{x}^{2}+\hat{\xi}^{2})
 \ln^{2}\!\left(\frac{\hat{x}}{\hat{x}-1}\right)
\nonumber\\
&\quad
-2\hat{\xi}(\hat{\xi}^{2}-1)
 \ln\!\left(\frac{\hat{x}}{1+\hat{x}}\right)
+\hat{\xi}(-1-2\hat{x}-2\hat{x}^{2}+\hat{\xi}^{2})
 \ln^{2}\!\left(\frac{\hat{x}}{1+\hat{x}}\right)
\nonumber\\
&\quad
+2(\hat{x}+\hat{\xi}-\hat{x}\hat{\xi}^{2}+\hat{\xi}^{3})
 \ln\!\left(\frac{\hat{x}}{\hat{x}-\hat{\xi}}\right)
+(-\hat{x}+\hat{\xi}+2\hat{x}^{2}\hat{\xi}
-\hat{x}\hat{\xi}^{2}-\hat{\xi}^{3})
 \ln^{2}\!\left(\frac{\hat{x}}{\hat{x}-\hat{\xi}}\right)
\nonumber\\
&\quad
+4\hat{\xi}
 \ln\!\left(\frac{\hat{x}-\hat{\xi}}{\hat{x}-1}\right)
+2(-\hat{x}-\hat{\xi}+\hat{x}\hat{\xi}^{2}+\hat{\xi}^{3})
 \ln\!\left(\frac{\hat{x}}{\hat{x}+\hat{\xi}}\right)
+(\hat{x}+\hat{\xi}+2\hat{x}^{2}\hat{\xi}
+\hat{x}\hat{\xi}^{2}-\hat{\xi}^{3})
 \ln^{2}\!\left(\frac{\hat{x}}{\hat{x}+\hat{\xi}}\right)
\Bigg\}.
\label{eq:unpolarized-DR-remainder}
\end{align}
For $t \neq 0$, the finite term in the notation of Ref.~\cite{Bhattacharya:2023wvy} is:
\begin{equation}
\begin{aligned}
\mathcal T_{\rm rem}^{S,t}(\hat{x},\hat{\xi})
\sim \frac{1}{(1-\hat{\xi}^{\,2})^2}
\Biggl\{&
(1-2\hat{x}+2\hat{x}^{\,2}-\hat{\xi}^{\,2})
\left[
 -
  \ln\!\frac{\hat{x}-1}{\hat{x}}
 +\frac{1}{2}\ln^2\!\frac{\hat{x}-1}{\hat{x}}
\right]
\\
&-\frac{(\hat{x}-\hat{\xi})
         (1-2\hat{x}\hat{\xi}-\hat{\xi}^{\,2})}{\hat{\xi}}
 \ln\!\frac{\hat{x}-\hat{\xi}}{\hat{x}}
 +\frac{\hat{x}(1+\hat{\xi}^{\,2})}{2\xi}
 \ln^2\!\frac{\hat{x}-\hat{\xi}}{\hat{x}}
-\frac{1+2\hat{x}^{\,2}-\hat{\xi}^{\,2}}{2}
 \ln\!\frac{\hat{x}-\hat{\xi}}{\hat{x}}\,
 \ln\!\frac{\hat{x}+\hat{\xi}}{\hat{x}}
\\
&+\frac{\hat{x}(1+\hat{\xi}^{\,2})}{\hat{\xi}}
 \operatorname{Li}_2\!\left(
 \frac{-2\hat{\xi}}{\hat{x}-\hat{\xi}}\right)
+(1-2\hat{x}+2\hat{x}^{\,2}-\hat{\xi}^{\,2})
 \left[
 \operatorname{Li}_2\!\left(\frac{1-\hat{\xi}}{1-\hat{x}}\right)
 +\operatorname{Li}_2\!\left(\frac{1+\hat{\xi}}{1-\hat{x}}\right)
 \right]
+(\hat{x}\to-\hat{x})\Biggr\}.
\end{aligned}
\label{eq:t-regulated-unpolarized-finite}
\end{equation}
Finally, moving onto the $m \neq 0$ case, the finite terms read:
\begin{align}
\mathcal T_{\rm rem}^{S,m}(\hat{x},\hat{\xi})
\sim{}&
-\frac{4}
{9\hat{\xi}(\hat{\xi}^{2}-1)}
\Bigg\{
-52\hat{\xi}+52\hat{\xi}^{3}
\nonumber\\
&\quad
-9\hat{\xi}
(-1+2\hat{x}-2\hat{x}^{2}+\hat{\xi}^{2})
\ln^{2}(\hat{x}-1)
+6\hat{\xi}\ln(\hat{x}-1)
\Bigl[
2(\hat{x}-1)^2
+(-3+6\hat{x}-6\hat{x}^{2}+3\hat{\xi}^{2})
 \ln\hat{x}
\Bigr]
\nonumber\\
&\quad
+12\hat{\xi}(1+2\hat{x}+\hat{x}^{2})\ln(1+\hat{x})
+9\hat{\xi}(1+2\hat{x}+2\hat{x}^{2}-\hat{\xi}^{2})
 \ln^{2}(1+\hat{x})
\nonumber\\
&\quad
+(-12\hat{x}^{2}\hat{\xi}
+24\hat{x}\hat{\xi}^{2}-12\hat{\xi}^{3})
 \ln(\hat{x}-\hat{\xi})
+(9\hat{x}-9\hat{\xi}-18\hat{x}^{2}\hat{\xi}
+9\hat{x}\hat{\xi}^{2}+9\hat{\xi}^{3})
 \ln^{2}(\hat{x}-\hat{\xi})
\nonumber\\
&\quad
+(-12\hat{x}^{2}\hat{\xi}
-24\hat{x}\hat{\xi}^{2}-12\hat{\xi}^{3})
 \ln(\hat{x}+\hat{\xi})
+(-9\hat{x}-9\hat{\xi}-18\hat{x}^{2}\hat{\xi}
-9\hat{x}\hat{\xi}^{2}+9\hat{\xi}^{3})
 \ln^{2}(\hat{x}+\hat{\xi})
\nonumber\\
&\quad
+6\ln\hat{x}\Bigl[
-4\hat{\xi}+4\hat{\xi}^{3}
+3\hat{\xi}(-1-2\hat{x}-2\hat{x}^{2}+\hat{\xi}^{2})
 \ln(1+\hat{x})
-3(\hat{x}-\hat{\xi}-2\hat{x}^{2}\hat{\xi}
+\hat{x}\hat{\xi}^{2}+\hat{\xi}^{3})
 \ln(\hat{x}-\hat{\xi})
\nonumber\\
&\qquad\qquad
+3\hat{x}\ln(\hat{x}+\hat{\xi})
+3\hat{\xi}\ln(\hat{x}+\hat{\xi})
+6\hat{x}^{2}\hat{\xi}\ln(\hat{x}+\hat{\xi})
+3\hat{x}\hat{\xi}^{2}\ln(\hat{x}+\hat{\xi})
-3\hat{\xi}^{3}\ln(\hat{x}+\hat{\xi})
\Bigr]
\Bigg\}.
\label{eq:symmetric-mass-remainder}
\end{align}

\section{Conclusions}
\label{sec:conclusions}
We have traced the anomaly-related poles in the gluon channel of
double deeply virtual Compton scattering (DDVCS) to a subleading Landau
singularity of the one-loop box amplitude. The singularity lies on
the boundary of the Feynman-parameter domain where the parameter
associated with the internal quark propagator joining the two
photon vertices vanishes. This propagator is inactive in the
Landau equations, and the other three form a triangle subdiagram.
The triangle (see Fig.~\ref{fig:z-zero-reduction}) thus identifies the specific boundary configuration
of the four-point integral responsible for the poles.

We identify the same pole-producing boundary of the one-loop box
across three infrared prescriptions: dimensional regularization
(DR), nonzero momentum transfer (\(t\neq0\)), and finite quark mass
(\(m\neq0\)). At \(t=m=0\), the box endpoint has a power (linear) infrared
divergence in four dimensions. By analytic continuation, DR conceals this power divergence,
which produces no \(1/\epsilon\) pole in the expansion about
\(d=4\). If evaluated in six dimensions,
the same endpoint would diverge logarithmically, giving a pole
in the expansion about \(d=6\).
Nonzero \(t\) and finite \(m\) expose this boundary contribution
through the known \(1/t\) anomaly-related pole and a previously
unreported \(1/m^2\) behavior, respectively. After summing the
contributing diagrams, we find that the two inverse-power terms
carry the same anomaly-related residue. Table~I summarizes the
boundary scaling for all three infrared prescriptions. These
results establish the triangle boundary of the one-loop box as
the common geometric origin of the anomaly-related poles. 

\begin{table}[!htbp]
\centering
\setlength{\cellspacetoplimit}{4pt}
\setlength{\cellspacebottomlimit}{0pt}
\begin{tabular}{c|S{c}|c}
Regulator & Representative boundary contribution & Singular behavior
\\[0.1cm] \hline
Cutoff or DR
&
\(\displaystyle
\int_{\delta}^{z_0}\frac{dz}{z^2}
\xrightarrow{\text{cutoff}}
\frac{1}{\delta};
\quad
\int_0^{z_0}\frac{dz}{z^{2+\epsilon}}
\xrightarrow[\epsilon\to0]{\text{analytic continuation}}
-\frac{1}{z_0}
\quad \text{(finite at \(d=4\))}
\)
&
Linear divergence (concealed by DR)
\\[4mm]
\(t\neq0\)
&
\(\displaystyle
\int_0^{z_0}
\frac{dz}{[a(x,y)z+b(x,y)t]^2}
\sim \int_0^{z_0}\frac{dz}{z^2}
\sim \frac{\mathcal R}{t}
\)
&
Inverse-momentum-transfer divergence
\\[4mm]
\(\ m\neq0\)
&
\(\displaystyle
\int_0^{z_0}
\frac{dz}{[a(x,y)z+m^2]^2}
\sim \int_0^{z_0}\frac{dz}{z^2}
\sim \frac{\mathcal R}{m^2}
\)
&
Inverse-mass-squared divergence
\end{tabular}
\caption{Schematic behavior of the four-point coefficient functions near
the triangle boundary \(z=0\), where \(z\) is the Feynman parameter
of the quark propagator joining the two photon vertices and
\(z_0\) is the upper cutoff defining the near-boundary region.
After summing the contributing diagrams, the \(t\)- and
\(m\)-regulated residues contain the same anomaly operator
structure, with potentially different coefficient functions.}
\label{tab:boundary-scaling-summary}
\end{table}

We also show that the pole residue depends on how the external
gluon polarizations are treated. One can retain the polarization
vectors explicitly or contract the two gluon Lorentz indices of
the box amplitude with a transverse projector. If one uses the
transverse projector, the box numerator can be written in terms
of inverse propagators. These factors cancel the corresponding propagators in the
integrand, suppressing the pole residue associated with the
box boundary and leaving bubble and triangle integrals
(the latter do not generate the anomaly-related pole). Retaining explicit polarization vectors
preserves the tensor structures associated with the scalar and
pseudoscalar gluonic operators, allowing us to identify the
corresponding anomaly-related terms. This projection dependence
may help explain why some of these terms were absent in earlier
calculations.

We also obtained the complete one-loop results for polarized and
unpolarized DDVCS with each infrared prescription. After summing
the box, cat's-eye, and hourglass diagrams, we find that, within
each polarization channel, the same evolution kernel governs
the infrared pole in DR and the logarithms \(\ln(Q^2/(-t))\)
and \(\ln(Q^2/m^2)\). The recovery
of the same kernel with all three prescriptions provides a
nontrivial consistency check of our one-loop results.

In conclusion, we have identified the triangle boundary of the
one-loop box as the common geometric origin of the anomaly-related
poles. We leave a systematic study of the joint \(t,m\to0\) limit,
its relation to DR, and the analytic
continuation of the associated loop functions to future work.
Our results provide a starting
point for identifying analogous poles in other hard-exclusive
processes.

\section*{Acknowledgment}
S.~B. thanks Yoshitaka Hatta for numerous helpful suggestions that improved the presentation of this manuscript. S.~B. also gratefully acknowledges stimulating discussions with Yoshitaka Hatta, Zhong-Bo Kang, Dmitri Kharzeev, Raju Venugopalan, and Werner Vogelsang.
This work was supported by the U.S. National Science Foundation under Grant No.~PHY-2609761. A.~Y. gratefully acknowledges support from the Mark Miller Award of the Department of Physics at the University of Connecticut, which made this undergraduate research contribution possible.

\appendix
\section{Passarino--Veltman functions}
\label{app:PV-notation}
Passarino--Veltman reduction expresses a one-loop tensor integral as a
sum of Lorentz tensors multiplied by scalar coefficient functions. The
letters \(A\), \(B\), \(C\), and \(D\) label integrals with one, two,
three, and four propagators, respectively. We use a two-point
(\(B\)-function) integral to illustrate how the tensor coefficients
and their indices are defined.

\begin{figure}[t]
\centering
\begin{tikzpicture}[line width=0.9pt, >=stealth]
  \coordinate (L) at (-1.1,0);
  \coordinate (R) at ( 1.1,0);

  \draw[->] (-2.1,0) -- (L) node[midway,above] {\(p\)};
  \draw[->] (R) -- (2.1,0) node[midway,above] {\(p\)};

  \draw (L) .. controls (-0.65,1.05) and (0.65,1.05) .. (R);
  \draw (L) .. controls (-0.65,-1.05) and (0.65,-1.05) .. (R);

  \fill (L) circle (2pt);
  \fill (R) circle (2pt);

  \node[above] at (0,0.82) {\(D_0,\ k\)};
  \node[below] at (0,-0.82) {\(D_1,\ k+p\)};
\end{tikzpicture}
\caption{Two-point, or bubble, topology defining the \(B\)-function
family. The external momentum satisfies \(p^2=s\); the two internal
propagators have masses \(m_0\) and \(m_1\).}
\label{fig:B-bubble}
\end{figure}
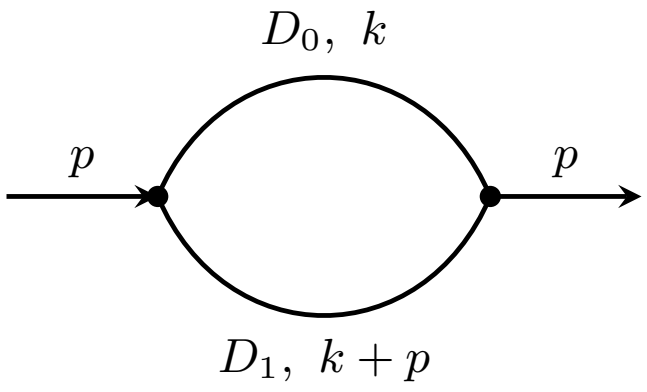

For the routing in Fig.~\ref{fig:B-bubble}, define
\begin{equation}
D_0=k^2-m_0^2+i0,
\qquad
D_1=(k+p)^2-m_1^2+i0.
\label{eq:B-propagators}
\end{equation}
Here \(i0\) denotes \(i\varepsilon\) with \(\varepsilon\to0^+\),
using the same causal prescription as in the Feynman-parameter
representations. The scalar \(B\)-function is the loop integral with numerator \(1\)
and denominator \(D_0D_1\). Tensor \(B\)-functions have factors of
the loop momentum \(k^\mu\) in the numerator. We write the scalar
and tensor integrals as
\begin{align}
B_0(s;m_0,m_1)
&\equiv
\int_k \frac{1}{D_0D_1},
\\
B^\mu(p)
&\equiv
\int_k \frac{k^\mu}{D_0D_1},
\\
B^{\mu\nu}(p)
&\equiv
\int_k \frac{k^\mu k^\nu}{D_0D_1}.
\label{eq:B-integrals}
\end{align}
Here \(\int_k\) denotes the dimensionally regulated loop integration
with the normalization used in Eq.~(5).
Since \(p^\mu\) is the only external vector in a two-point integral,
Lorentz covariance gives
\begin{align}
B^\mu(p) &= p^\mu B_1(s;m_0,m_1),
\\
B^{\mu\nu}(p)
&=g^{\mu\nu}B_{00}(s;m_0,m_1)
  +p^\mu p^\nu B_{11}(s;m_0,m_1).
\label{eq:B-tensor-decomposition}
\end{align}
Thus, \(B_1\) is the scalar coefficient multiplying one external
momentum. At rank two, \(B_{00}\) multiplies the metric tensor, while
\(B_{11}\) multiplies two external momenta. The scalar function
\(B_0\) has no Lorentz indices.
Thus, the subscripts record these Lorentz structures. In the notation of
Eq.~(2), each pair of zero indices represents one metric tensor, and
each index \(1\) represents one factor of \(p^\mu\). The following
examples make the counting explicit:
\begin{center}
\begin{tabular}{c c c c}
\hline
\(r\) & \(n_1\) & Coefficient & Tensor structure \\
\hline
\(0\) & \(0\) & \(B_0\) & scalar \\
\(0\) & \(1\) & \(B_1\) & \(p^\mu B_1\) \\
\(1\) & \(0\) & \(B_{00}\) & \(g^{\mu\nu}B_{00}\) \\
\(0\) & \(2\) & \(B_{11}\) & \(p^\mu p^\nu B_{11}\) \\
\hline
\end{tabular}
\end{center}
Here \(r\) counts metric tensors and \(n_1\) counts factors of the
external momentum. The tensor rank is \(2r+n_1\). 

The parameter representation follows by combining the two
propagators:
\begin{equation}
\frac{1}{D_0D_1}
=
\int_0^1\mathrm{d}x\,
\frac{1}{
\bigl[(k+xp)^2-F_B(x)+i0\bigr]^2},
\end{equation}
where
\begin{equation}
F_B(x)
=sx^2+(-s+m_1^2-m_0^2)x+m_0^2.
\label{eq:FB-appendix}
\end{equation}
After shifting the loop momentum to \(q=k+xp\), the rank-one
numerator becomes
\begin{equation}
k^\mu=q^\mu-xp^\mu.
\end{equation}
The term proportional to \(q^\mu\) vanishes upon integration because
it is odd in the shifted loop momentum. The remaining term is
proportional to \(-x p^\mu\). This explains why the parameter
representation for \(B_1\) contains one power of \(x\), with a minus
sign relative to \(B_0\). At rank two, the numerator contains
\[
k^\mu k^\nu
=q^\mu q^\nu-x(q^\mu p^\nu+p^\mu q^\nu)
  +x^2p^\mu p^\nu.
\]
The terms linear in \(q^\mu\) again vanish. The \(q^\mu q^\nu\) term
generates the metric contribution \(B_{00}\), while the
\(x^2p^\mu p^\nu\) term generates \(B_{11}\).
For completeness, the resulting coefficients in the normalization
of Eq.~(5) are
\begin{align}
B_0(s;m_0,m_1)
&=
\mu^{2\epsilon}e^{\gamma_E\epsilon}\,
\Gamma(\epsilon)
\int_0^1\mathrm{d}x\,
\bigl[F_B(x)-i\varepsilon\bigr]^{-\epsilon},
\nonumber\\
B_1(s;m_0,m_1)
&=
-\mu^{2\epsilon}e^{\gamma_E\epsilon}\,
\Gamma(\epsilon)
\int_0^1\mathrm{d}x\,x\,
\bigl[F_B(x)-i\varepsilon\bigr]^{-\epsilon},
\nonumber\\
B_{00}(s;m_0,m_1)
&=
-\frac{\mu^{2\epsilon}e^{\gamma_E\epsilon}}{2}\,
\Gamma(\epsilon-1)
\int_0^1\mathrm{d}x\,
\bigl[F_B(x)-i\varepsilon\bigr]^{1-\epsilon},
\nonumber\\
B_{11}(s;m_0,m_1)
&=
\mu^{2\epsilon}e^{\gamma_E\epsilon}\,
\Gamma(\epsilon)
\int_0^1\mathrm{d}x\,x^2\,
\bigl[F_B(x)-i\varepsilon\bigr]^{-\epsilon}.
\label{eq:B-coefficients-worked}
\end{align}
These are the cases \(r=0,n_1=0\), \(r=0,n_1=1\),
\(r=1,n_1=0\), and \(r=0,n_1=2\), respectively, of the general
two-point representation in Eq.~(5).

The same index-counting rule applies to the three- and four-point
functions. Their indices \(1,2,3\) label the independent external
momenta used in the tensor decomposition. For example,
\(D_{00123}\) contains one pair of zero indices, corresponding to one
metric tensor, and one factor of each of the momenta labelled
\(1,2,3\). Its tensor rank is therefore five. In the parameter
representation of Eq.~(10), the momentum labels \(1,2,3\) correspond
to the factors \(x,y,z\), respectively, so this coefficient carries
the numerator factor \(xyz\). 

\bibliography{references.bib}

\end{document}